\documentclass[twocolumn,twoside,reprint,superscriptaddress,aps,floatfix]{revtex4-1}
\usepackage{physics}
\usepackage{verbatim}
\usepackage{graphicx}
\usepackage[usenames]{color}
\usepackage{microtype}
\usepackage{amsmath}
\usepackage{amssymb}
\usepackage{xspace}
\usepackage{footnote}
\usepackage{relsize}

\newcommand{\xj}{\mathcal{X}_{j}}
\newcommand{\xk}{\mathcal{X}_{k}}

\newcommand{\calx}{\mathcal{X}}

\newcommand{\CA}{\mathcal{A}}
\newcommand{\loss}{\ell^2}

\newcommand{\balpha}{\boldsymbol{\alpha}}
\newcommand{\bmu}{\boldsymbol{\mu}}
\begin{document}

\title{
Accurate molecular polarizabilities with coupled-cluster theory and machine learning 
}

\newcommand{\red}[1]{{\color{red} #1}}

\author{David M. Wilkins}
\affiliation{Laboratory of Computational Science and Modeling, IMX, \'Ecole Polytechnique F\'ed\'erale de Lausanne, 1015 Lausanne, Switzerland}

\author{Andrea Grisafi} 
\affiliation{Laboratory of Computational Science and Modeling, IMX, \'Ecole Polytechnique F\'ed\'erale de Lausanne, 1015 Lausanne, Switzerland}

\author{Yang Yang}
\affiliation{Department of Chemistry and Chemical Biology, Cornell University, Ithaca, NY 14853, USA}

\author{Ka Un Lao}
\affiliation{Department of Chemistry and Chemical Biology, Cornell University, Ithaca, NY 14853, USA}

\author{Robert A. DiStasio Jr.}
\affiliation{Department of Chemistry and Chemical Biology, Cornell University, Ithaca, NY 14853, USA}

\author{Michele Ceriotti}
\email{michele.ceriotti@epfl.ch}
\affiliation{Laboratory of Computational Science and Modeling, IMX, \'Ecole Polytechnique F\'ed\'erale de Lausanne, 1015 Lausanne, Switzerland}

\begin{abstract}

The molecular polarizability describes the tendency of a molecule to deform or polarize in response to an applied electric field. As such, this quantity governs key intra- and inter-molecular interactions such as induction and dispersion, plays a key role in determining the spectroscopic signatures of molecules, and is an essential ingredient in polarizable force fields and other empirical models for collective interactions. 
Compared to other ground-state properties, an accurate and reliable prediction of the molecular polarizability is considerably more difficult as this response quantity is quite sensitive to the description of the underlying molecular electronic structure. 
In this work, we present state-of-the-art quantum mechanical calculations of the static dipole polarizability tensors of 7,211 small organic molecules computed using linear-response coupled-cluster singles and doubles theory (LR-CCSD). Using a symmetry-adapted machine-learning based approach, we demonstrate that it is possible to predict the molecular polarizability with LR-CCSD accuracy at a negligible computational cost. The employed model is quite robust and transferable, yielding molecular polarizabilities for a diverse set of 52 larger molecules (which includes challenging conjugated systems, carbohydrates, small drugs, amino acids, nucleobases, and hydrocarbon isomers) at an accuracy that exceeds that of hybrid density functional theory (DFT). The atom-centered decomposition implicit in our machine-learning approach offers some insight into the shortcomings of DFT in the prediction of this fundamental quantity of interest.
\end{abstract}

\maketitle

\section{Introduction}
The last decade has seen great progress in the first-principles evaluation of the stability and properties of materials and molecules. 
Kohn-Sham density functional theory (DFT) has played a pivotal role in this endeavor by providing ground-state properties with an accuracy that is sufficient for many useful applications at a manageable computational cost~\cite{engel2011,burke2012,cottenier2016}.
However, DFT is not equally accurate for every property of interest. For instance, an accurate and reliable description of the molecular dipole polarizability $\balpha$, a tensor which describes the tendency of a molecule to deform (or polarize) in the presence of an applied electric field $\mathbf{E}$, can be quite difficult to obtain. This is primarily due to the fact that $\balpha$ is a response property that is particularly sensitive to the quantum mechanical description of the underlying electronic structure. As such, non-trivial electron correlation effects and basis set incompleteness error must be simultaneously accounted for when determining $\balpha$~\cite{stone1997theory,doi:10.1021/acs.chemrev.6b00446}. %
For these reasons, and in light of the fact that $\balpha$ is a fundamental quantity of interest that underlies induction and dispersion interactions~\cite{tkat-sche09prl,tkatchenko2012,grimme2011,grimme_chapter}, %
Raman and sum frequency generation (SFG) spectroscopy~\cite{Luber2014,Zhang2016,Medders2016,Morita2000,Morita2002,shen1989}, and represents a key ingredient in the development of next-generation polarizable force fields~\cite{Sprik1988,Fanourgakis2008,Lopes2009,pond+10jpcb,medd+14jctc,Cisneros2016}, it is important to provide benchmark values for $\balpha$ beyond the accuracy of DFT. 
In this regard, linear-response coupled-cluster theory including single and double excitations (LR-CCSD) has been shown to provide considerably more accurate and reliable predictions for $\balpha$ when used in conjunction with a sufficiently large (diffuse) one-particle basis set~\cite{hait_head-gordon_2018,doi:10.1063/1.2929840,LaoJiaGla18}. However, such a prediction is accompanied by a substantially larger computational cost (scaling with the sixth power of the system size), which becomes quite prohibitive when treating even moderate-size molecules with around $15$ atoms. %

In the last few years, machine learning has gained traction as an alternative approach to the prediction of molecular properties, substituting or complementing electronic-structure methods~\cite{behl-parr07prl,bart+10prl,rupp+12prl}.
In particular, it has been shown that accuracy on par with (or even better than) DFT can be achieved in the prediction of many molecular properties~\cite{de+16pccp, fabe+17jctc}, and that DFT~\cite{rama+15jctc} or coupled-cluster~\cite{bart+17sa} accuracy can be reached more easily when using a more computationally efficient (and quite often less accurate) electronic structure method as a stepping stone.
The polarizability, however, poses an additional challenge to machine-learning, due to its tensorial nature, and the predicted $\balpha$ must transform according to the symmetries of the $SO(3)$ rotation group. 
For rigid molecules, this is easily achieved by learning the components of the tensor written in the reference frame of the molecule~\cite{tris+15jctc,lian+17prb}. However, to obtain a transferable model that would also be suitable for flexible molecules--as well as different compounds--this line of thought would require a cumbersome and inelegant fragment decomposition. 
To avoid these complications, a recently developed Gaussian process regression scheme adapted to rotational symmetry (Symmetry-Adapted GPR or SA-GPR), has been derived to naturally incorporate this $SO(3)$ covariance into a ML scheme that is suitable to predict tensorial quantities of arbitrary order~\cite{gris+18prl}. 
In this paper, we present a comprehensive benchmark of molecular polarizabilities at the coupled-cluster level, based on the QM7b database which is composed of more than 7,000 small molecules~\cite{mont+13njp}. %
We then use these reference calculations to assess the accuracy of different hybrid DFT schemes, and use a SA-GPR scheme to train on QM7b a ML model (ALPHA-ML) that can inexpensively predict the polarizability tensor. 
We then test the extrapolative prediction capabilities of ALPHA-ML on a showcase dataset composed of 52 larger molecules, and demonstrate that this approach provides a viable alternative to state-of-the-art electronic structure methods for predicting the polarizability of molecules.

\section*{Results}
\newcommand{\frob}[1]{\ensuremath{\left\|#1\right\|_F}}

\subsection*{Electronic Structure Calculations}

In this work, we utilized the QM7b database~\cite{blum,rupp+12prl,mont+13njp}, which contains 7,211 molecules containing up to 7 ``heavy'' atoms (\textit{i.e.}, C, N, O, S, and Cl) with varying levels of H saturation. This database is based on a systematic enumeration of organic compounds~\cite{blum}, and  contains a rich diversity of molecular structures (including double and triple bonds, (hetero)cycles, carboxyl, cyanide, amide, alcohol, and epoxy groups), and provides a challenging test of the accuracy associated with DFT and quantum chemical methodologies. DFT-based molecular polarizabilities were obtained by (numerical) differentiation of the molecular dipole moment, $\bmu$, with respect to an external electric field $\mathbf{E}$, \textit{i.e.}, $\balpha = \partial \mathbf{\bmu} / \partial \mathbf{E}$, using the hybrid B3LYP~\cite{doi:10.1063/1.464913,doi:10.1021/j100096a001} and SCAN0~\cite{doi:10.1063/1.4940734} functionals.
Reference molecular polarizabilities were obtained using LR-CCSD~\cite{doi:10.1063/1.458814,doi:10.1002/(SICI)1097-461X(1998)68:1<1::AID-QUA1>3.0.CO;2-Z}. To combat the basis set incompleteness error, which can be even more important than higher-order (\textit{i.e.}, beyond single and double excitations) electron correlation effects in an accurate and reliable determination of $\balpha$~\cite{HamGovKow09,LaoJiaMai18,LaoJiaGla18}, we employed the d-aug-cc-pVDZ basis set~\cite{doi:10.1063/1.462569} for all calculations herein. 
Even though this double-$\zeta$ basis set has only a moderate number of polarization functions, augmentation with two sets of diffuse functions aids tremendously in the convergence of $\balpha$~\cite{ChrGauSta99,HamGovKow09,LaoJiaMai18,LaoJiaGla18}. For a more detailed description of the electronic structure calculations performed in this work, see the Materials and Methods section. 

For the QM7b database, B3LYP predicts $\balpha$ with a MSE of $0.259$~a.u., a MAE of $0.302$~a.u., and a RMSE of $0.404$~a.u. with respect to the reference LR-CCSD values. To enable comparisons between molecules of different sizes, all the error estimates (explicit expressions for which are given in the Materials and Methods section) are computed based on the molecular polarizabilites scaled by the number of atoms $n_i$ contained within a given molecule. These errors, which include both scalar and anisotropic contributions, are quite substantial and correspond to nearly $20\%$ of the intrinsic variability within the database, defined as $\sigma_\text{CCSD} = \sqrt{\frac{1}{N}
\sum_i \frac{1}{n_i^2}\frob{\balpha^{(\text{CCSD})}_i-\left<\balpha^{(\text{CCSD})}\right>}^2}$ . %
The large MSE value indicates a systematic overestimation of $\balpha$ by B3LYP~\cite{KARNE2015168,hait_head-gordon_2018}.
The results from the hybrid meta-GGA functional SCAN0 show a substantially reduced MSE of $0.059$~a.u. Despite the smaller  systematic overestimation of $\balpha$ in comparison with B3LYP, the SCAN0 statistical errors are still very large, with computed MAE and RMSE values of $0.217$~a.u. and $0.316$~a.u., respectively. From the point of view of machine learning, our model performs almost equally well for B3LYP and SCAN0. For this reason, we will focus our discussion on the B3LYP and LR-CCSD results, which will be referred to as DFT and CCSD, respectively, throughout the remainder of the manuscript.

\subsection*{Improved Symmetry-Adapted Gaussian Process Regression}

The formalism underlying the SA-GPR scheme in general and the $\lambda$-SOAP descriptors on which our model is based have been introduced elsewhere~\cite{gris+18prl} and are summarized in the Materials and Methods section.
In this work, however, we include several substantial improvements, which make the SA-GPR model more accurate and faster to compute, and are worth a separate discussion.

Evaluation of the $\lambda$-SOAP representation is greatly accelerated by sparsification: by taking a few hundred spherical harmonic components (out of several tens of thousands) using farthest-point sampling \cite{imba+18jcp}, the calculation of the kernel in  Eq.~\ref{eq:combine_power_spectra} can be carried out with essentially the same result as if all components were retained, but with a much lower computational cost.
A second improvement of the SA-GPR scheme is the generalization of the $\lambda$-SOAP kernels beyond the linear kernels used in Ref.~\citenum{gris+18prl}. 
It has been shown that in many cases taking an integer power of the scalar SOAP kernel improves the performance of the associated ML model. This can be understood in terms of the order (2-body, 3-body, \ldots)  of the interatomic correlations that are described by different kernels~\cite{bart+13prb,glie+18prb,density-arxiv}. 
In the tensorial case, one should be careful as the linear nature of the kernel is essential to ensure the correct covariant behavior. To include non-linearity and increase the body order of the model without affecting the symmetry properties, we multiplied the  $\lambda>0$ kernels by the scalar $\lambda=0$ kernel raised to the power of $\zeta-1$, as in Eq.~\ref{eq:nonlinear}.
Finally, we explored the possibility of combining multiple kernels computed with different environment radii $r_\text{c}$, which has been shown to improve the model accuracy in the scalar case~\cite{bart+17sa}. Together, these improvements double the model accuracy on QM7b, as discussed in detail in the Supplementary Information (SI).

\subsection*{Learning on the QM7b Database}\label{sec:QM7b}

These state-of-the-art reference calculations and SA-GPR scheme lay the foundations for a transferable model to predict molecular polarizabilities. We call this model ALPHA-ML, and in this first incarnation we use for training the reference DFT and CCSD calculations on the QM7b set~\cite{mont+13njp}.
As a first verification of the performance of the model, we compute learning curves on the CCSD polarizabilities of the QM7b dataset, using up to 5,400 structures for training and assessing the generalization power of the model by predicting the value of $\balpha$ for 1,811 structures that are not included in the training. 
\begin{figure}[h!]%
\centering
\includegraphics[width=\linewidth]{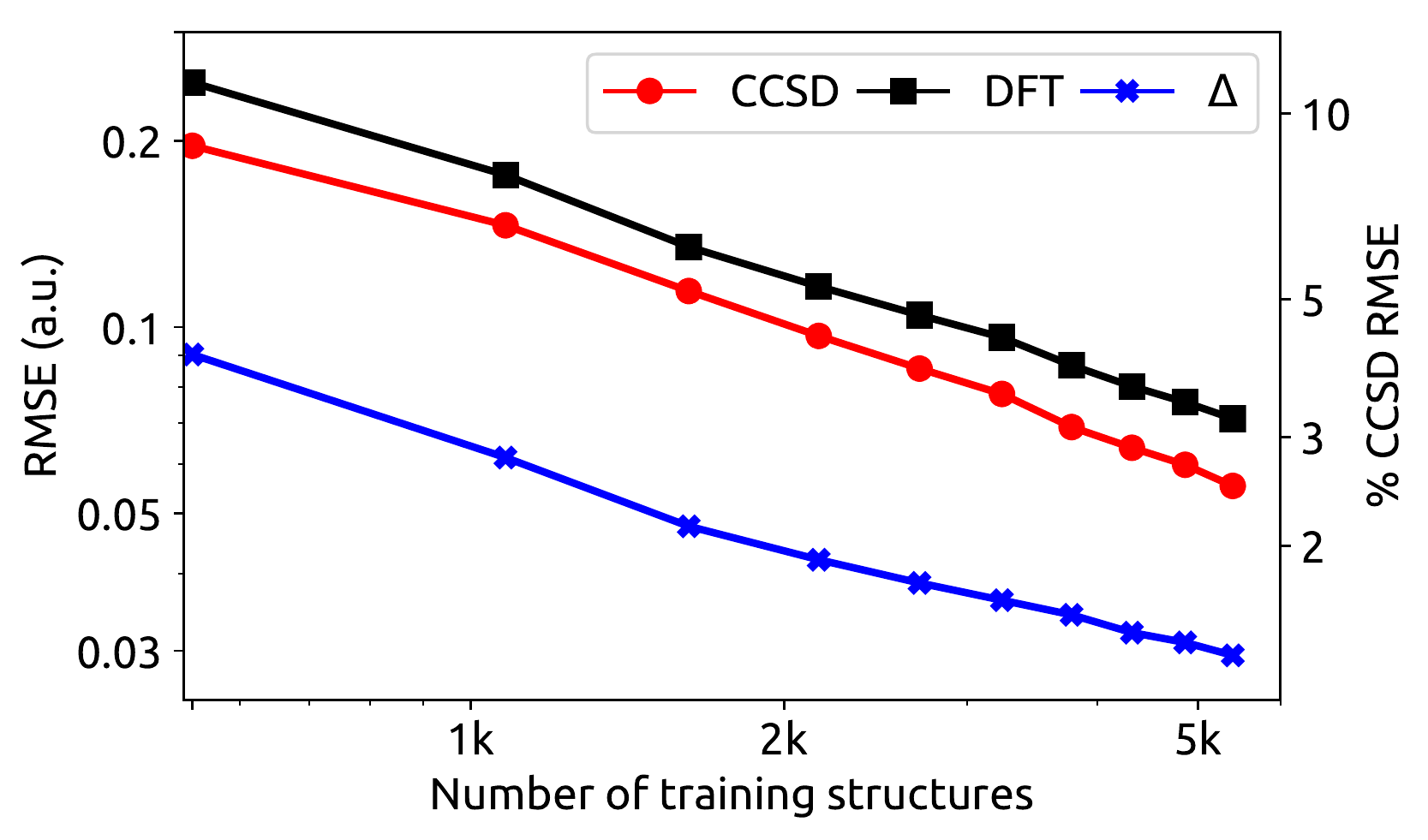}
\caption{Learning curves for the per-atom polarizabilities of the QM7b molecules,
calculated using either CCSD or DFT, as well as for the difference ($\Delta$) between the two.
The testing set consists of 1811 molecules, and the right-hand axis shows the RMSE as a fraction of the intrinsic variability of the CCSD polarizability, $\sigma_\text{CCSD}$.}\label{fig:ccsd_dft_delta_DADZ}
\end{figure}
\begin{figure*}[htb]
\centering
\includegraphics[width=0.98\linewidth]{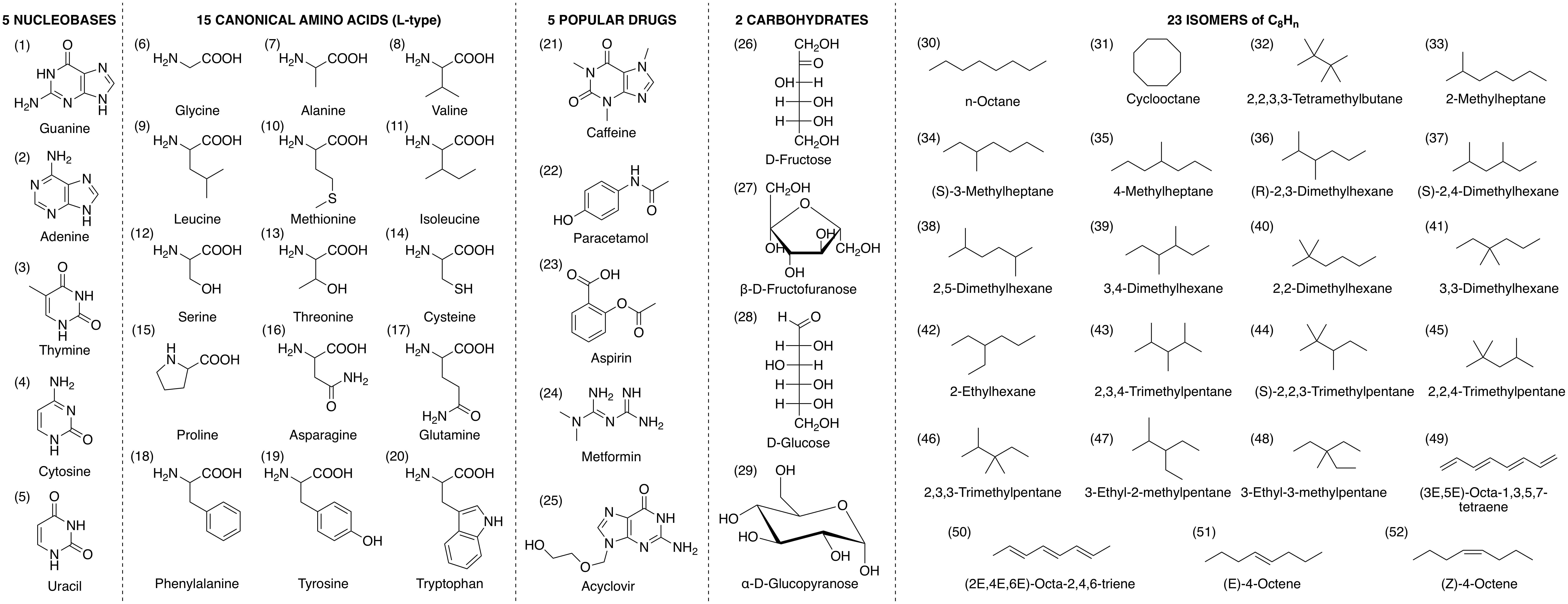}
\caption{List of molecules included in the showcase dataset. Numbers refer to the position in the dataset and are used for reference in other figures.}
\label{fig:all-show}
\end{figure*}

Using the best kernel hyperparameters (as described in the SI), we trained a model to learn the CCSD polarizability.
We report ML errors in terms of the percentage of the intrinsic variability of the CCSD dataset ($\sigma_\text{CCSD}=2.216$~a.u. per atom), so as to provide a direct measure of the learning performance. As illustrated by the learning curves in Fig.~\ref{fig:ccsd_dft_delta_DADZ}, using 75\% of the QM7b database for training gives a 2.5\% RMSE as a fraction of $\sigma_\text{CCSD}$ in predicting the CCSD polarizability.
 
To get a clearer idea of the accuracy of these predictions, one can compare with the accuracy of density functional theory. 
Using the same metric, the error of DFT relative to CCSD is 18\% of the standard deviation of the corresponding CCSD values. 
A machine-learning model based on SA-GPR can thus be trained to give polarizabilities with an order of magnitude accuracy better than DFT, relative to the CCSD reference values.
The DFT polarizability itself can be learned
with an error of 3.2\% of the $\sigma_\text{CCSD}$.
As seen in other cases~\cite{bart+17sa,rama+15jctc}, high-end quantum chemistry calculations are smoother, and slightly easier to learn, than more approximate electronic-structure methods.

The model can also be trained to evaluate corrections between different levels of theory. It is often the case that such corrections can be modelled with much smaller error than the raw quantity~\cite{rama+15jctc,bart+17sa}.
For instance, using DFT as a baseline to learn CCSD polarizabilities reduces the error by a factor of about 2 relative to the direct learning of $\balpha_\text{CCSD}$ (Fig.~\ref{fig:ccsd_dft_delta_DADZ}). This provides a way to further reduce the prediction error at the cost of performing a baseline DFT calculation. 
In the SI we demonstrate that the performance of ALPHA-ML is rather insensitive to the details of the reference method, showing similar accuracy for the SCAN0 functional as that observed for B3LYP.

\subsection*{Extrapolation to Larger Molecules}\label{sec:showcase}

Our definition of the kernel between two molecules as an average of environmental kernels means that the polarizabilities predicted by ALPHA-ML are given as a sum of predicted polarizabilities of each environment~\cite{bart+17sa}.
This feature of the model allows it to predict $\balpha$ for larger and more complex molecules. To test the behavior of the model in this extrapolative regime, we trained it on the full QM7b set and predicted the polarizabilities of a showcase set of 52 large molecules, including amino acids, nucleobases, drug molecules, carbohydrates, alkanes and alkenes (See Fig.~\ref{fig:all-show}, and the SI for full details).

\begin{table}[tbhp]
\centering
\caption{Root mean square errors in machine-learning of the per-atom polarizabilities of the  showcase molecules. CCSD/DFT denotes the discrepancy between CCSD and DFT calculations, while CCSD/ML and DFT/ML give the errors in predicting CCSD and DFT $\boldsymbol{\alpha}_{n}$ respectively, using a machine-learning model. $\Delta$(CCSD-DFT)/ML gives the error in predicting the difference between CCSD and DFT polarizability. In all cases, the full QM7b Database is used as a training set.
The total RMSE, expressed in atomic units per atom, is broken down into the errors associated with the scalar ($\lambda=0$) and tensorial ($\lambda=2$) components.
}\label{tab:showcase_errors}
\begin{tabular}{cccc}
\hline\hline
Method & RMSE & RMSE($\lambda=0$) & RMSE($\lambda=2$)\\
\hline
CCSD/DFT & 0.573 & 0.348  & 0.456  \\
CCSD/ML & 0.304 & 0.120 & 0.285 \\
DFT/ML & 0.403 & 0.143 & 0.377  \\
$\Delta$(CCSD-DFT)/ML & 0.212 & 0.083  & 0.196 \\
\hline\hline
\end{tabular}
\end{table}

In Table~\ref{tab:showcase_errors} we show the RMSE errors in learning  $\balpha$ for the showcase molecules, either using CCSD or DFT, as well as the error made when using the DFT results as an approximation to CCSD.
We also break down the error into the $\lambda=0$ and $\lambda=2$ components of the polarizability, which shows that the error in the anisotropic response is comparable to that in the trace, and that ALPHA-ML learns both components with similar efficiency.
As in the previous section we note that using an ALPHA-ML model to predict the CCSD polarizability is more accurate than using DFT, and using DFT as a baseline leads to a further reduction of $\sim$20-30\%{}. 
In the SI we discuss further the behavior of the model when using the SCAN0 functional, which is similar to that observed here for B3LYP. 
While ALPHA-ML predicts CCSD polarizabilities of the showcase molecules with better-than-DFT accuracy, the overall performance is considerably worse than that seen for the validation set in the QM7b database. 

\begin{figure}[htb]
\centering
\includegraphics[width=\linewidth]{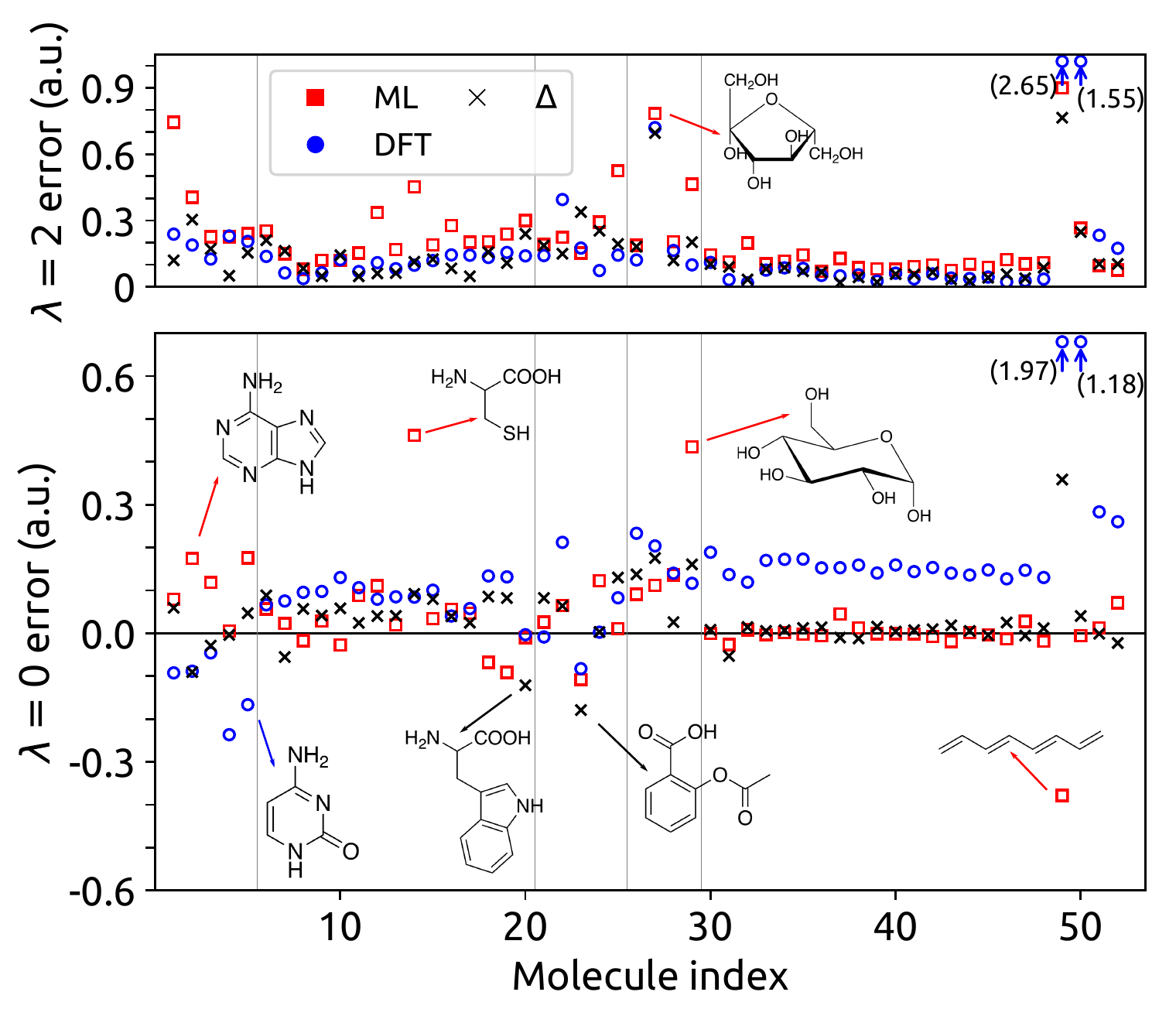}
\caption{
Error made in approximating the $\lambda=0$ (bottom panel) and $\lambda=2$ (top panel) components of the average polarizability per atom for the 52 showcase molecules, as a function of the molecule indices in Fig.~\ref{fig:all-show}. Vertical lines show the partitioning of these molecules into different groups.
Red squares show the machine-learning error, blue circles the error made in using the DFT polarizability to approximate the CCSD polarizability, and black crosses the error when $\Delta$-learning of the correction to the DFT polarizability is used.
\label{fig:showcase_parity_plot_ccsd_vs_dft}
}
\end{figure}

We can investigate the performance of the model in more detail by analyzing the errors of individual molecules in the showcase set. 
Fig.~\ref{fig:showcase_parity_plot_ccsd_vs_dft} shows that errors are very small for most molecules.
Large errors occur predominantly for highly-polarizable compounds, particularly those that show a large degree of conjugation, such as long-chain alkenes and purine nucleobases. For these systems, the electronic structure is characterized by a high degree of delocalization,
requiring larger cutoffs and more complex reference molecules to obtain accurate predictions.
The ML predictions for the tensorial component of the polarizability, $\balpha^{(2)}$,
tend to be slightly less accurate than the DFT reference, except for the highly-polarizable alkenes, for which the model outperforms DFT dramatically. 
Sulfur-containing structures, which are poorly represented in QM7b, also exhibit comparatively large errors.
Interestingly, many of the problematic molecules also show a large discrepancy between CCSD and DFT, suggesting that the same collective behavior that complicates ML also leads to poor performance of approximate electronic structure methods.

\subsection*{Atom-Centered Environmental Polarizabilities} 

The atom-centered structure of ALPHA-ML provides a natural additive decomposition of $\balpha$ into a sum of local terms, $\sum_i\balpha_i$, which can be used to better understand how different functional groups contribute to the molecular polarizability.
Unlike other methods for decomposing the polarizability, such as an atoms-in-molecules scheme~\cite{laidig1990} or a self-consistent decomposition~\cite{applequist1972,DelloStritto2017}, the approach used in this section does not require any additional calculations on top of that of molecular polarizability, as the atom-centered polarizabilities come as a byproduct of the SA-GPR scheme, which is based on local environmental kernels.
When interpreting $\balpha_i$s, one should keep in mind that each term corresponds to the contribution from  \emph{the entire atom-centered environment}, and the way the polarizability is split between neighboring atoms is entirely inductive, reflecting the interplay between data, structure (as represented by the kernels) and regression rather than explicit physical considerations. 
For instance, a few atoms within the showcase set (in particular several H environments) have $\balpha_i$ with negative eigenvalues, which reflects how they contribute to reduce the dielectric response of the functional group they are part of. 

\begin{figure}[htb]
\centering
\includegraphics[width=\linewidth]{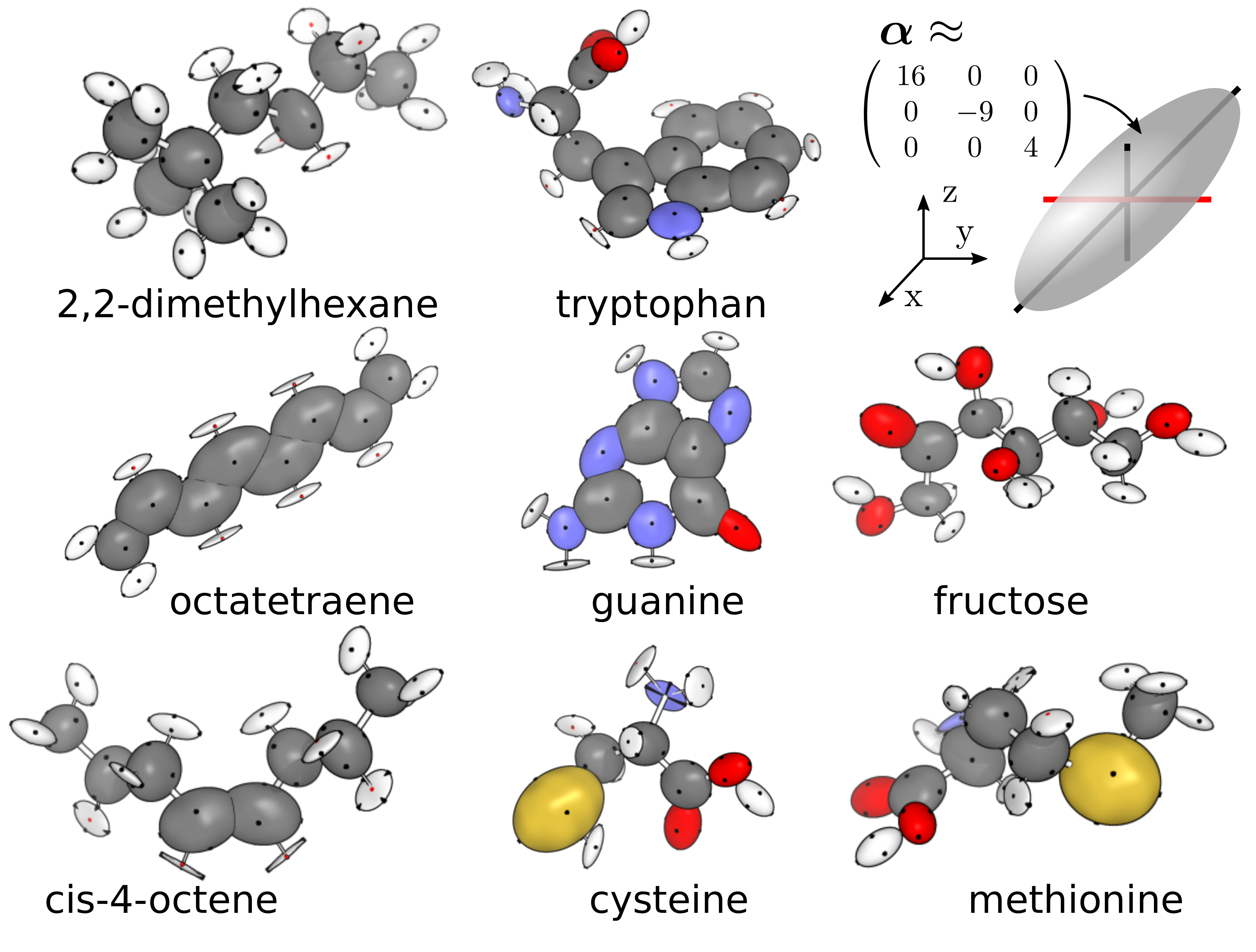}
\caption{Predicted atomic contributions to the total CCSD polarizability tensor for a selection of molecules in the showcase set. The ellipsoids are aligned along the principal axes of the atomic polarizability, and their extent is proportional to the square root of the corresponding eigenvalue of $\balpha_i$.
The ellipsoids have dimensions that are proportional to the modulus of the square root of eigenvalues of $\Delta\balpha_i$. The principal axes are shown, and are colored based on whether the corresponding eigenvalues are positive (black) or negative(red) See also the figure key, not to scale.
}\label{fig:anisotropies_showcase}
\end{figure}

With this in mind, one can clearly recognize physically-meaningful features in the magnitude and anisotropy of the $\balpha_i$. 
Consider the eight representative examples from the showcase dataset, depicted in Fig.~\ref{fig:anisotropies_showcase}.
Comparing saturated and unsaturated hydrocarbons (e.g. 2,3-dimethylhexane, cis-4-octene and octatetraene, in the figure) one sees that ALPHA-ML predicts that the contribution from unsaturated carbon atoms is both large and very anisotropic, consistent with the high electron delocalization along conjugated sections.
Similarly high and anisotropic contributions are also associated with aromatic systems, as seen for instance in the indole ring of tryptophan, and in adenine.
Oxygen atoms are generally assigned a very anisotropic $\balpha_i$; a large fraction of the polarizability of hydroxyl and carboxyl groups is assigned to the environments centered around nearby H and C atoms, but O atoms systematically contribute a further anisotropic term that is usually oriented perpendicular to the highly-polarizable lone pairs (see e.g. fructose, as well as the carboxyl group in the amino acids). 
Of all the atoms in the showcase set,  the sulfur-centered environments in cysteine and methionine have the largest contribution to the total polarizability, and exhibit a strongly anisotropic response.
The better-than-DFT performance of ALPHA-ML when faced with the challenge of molecules that are different and much larger than those included in the training set can then be understood based on the ability of the model to determine an atom-centered decomposition that assigns meaningful contributions to the total polarizability from functional groups, based on relatively local structural information.

\begin{figure}[tbp]
\centering
\includegraphics[width=\linewidth]{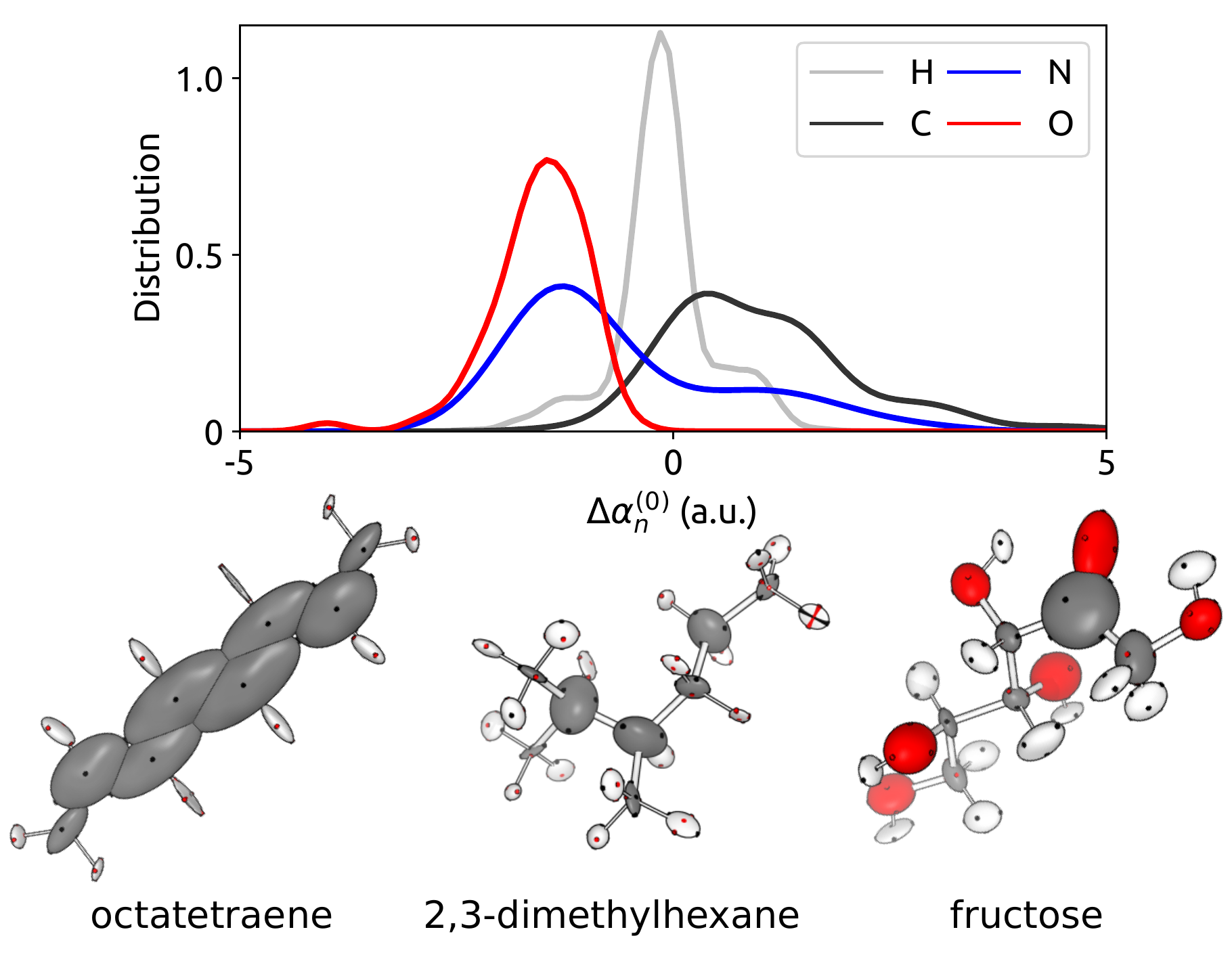}
\caption{Top: distributions of the predicted atomic contribution to the $\lambda=0$ component  of the difference between DFT and CCSD polarizability. Bottom: example decompositions of the polarizability difference. The ellipsoids represent the magnitude and principal axes of $\Delta\balpha_i$. Black axes indicate that DFT polarizability is larger than  CCSD, red axes that DFT polarizability is smaller. See figure key in Fig.~\ref{fig:anisotropies_showcase}. }\label{fig:delta-decomposition}
\end{figure}

An atom-centered ML model is even more useful when built on the difference between different reference electronic-structure methods, as it can help interpret the discrepancy between the different techniques in terms of contributions from specific functional groups.
As shown in Fig.~\ref{fig:delta-decomposition}, the $\Delta$-learning predictions between CCSD and DFT systematically attributes positive contributions to C atoms, and negative contributions to O atoms. 
This suggests that DFT tends to overestimate the polarizability of carbon-centered groups, and to underestimate the polarisability of oxygen-containing moieties. 
Inspecting individual $\Delta\balpha_i$ provides more detailed insight, identifiying known deficiencies of the approximate methods. For instance, DFT substantially overestimates $\balpha$ for conjugated systems, with relatively low errors for the saturated carbon atoms in dimethylhexane and fructose, and the large error on octatetraene being associated mostly with the delocalized $\pi$ electrons. 
The systematic underestimation of the contribution from oxygen-containing environments atoms by DFT, instead, appears to be relatively isotropic. 

\section*{Discussion}

Polarizability calculations with traditional methods have always implied a trade-off between accuracy and computational cost. 
CCSD calculations give more accurate prediction of the polarizability of molecules (especially large molecules) than DFT with various functionals~\cite{hait_head-gordon_2018,doi:10.1063/1.2929840,LaoJiaGla18},
but their computational cost can be prohibitive.
In our case, the LR-CCSD calculations of very large showcase molecules
took up to hundreds of hours and thousands of gigabytes of RAM.
The improvement in accuracy on going from DFT to CCSD comes with a steep increase in computational cost.

In this paper we have shown that the combination of SA-GPR with $\lambda$-SOAP kernels and CCSD reference calculations on small molecules that underlies the ALPHA-ML framework allows us to sidestep these expensive calculations, obtaining results with at least DFT accuracy -- and often much better -- with a fraction of the computational cost.
A model trained on a database of small molecules can be extrapolated to predict larger and more complex compounds, with a prediction quality that rivals DFT and can be systematically improved by extending the training set.
The atom-centered decomposition of the ML predictions of $\balpha$ can be interpreted in terms of physical-chemical considerations, revealing for instance the large, anisotropic contribution from delocalised $\pi$ systems, although in doing so one should keep in mind that the contributions are related to chemical environments rather than to a literal atoms-in-molecules kind of decomposition.  
A similar analysis performed on the difference between CCSD and DFT polarizabilities is also very useful to pinpoint the portions of a molecules and the functional groups that contribute the most to the polarizability error, guiding the development of better approximations. 

Having shown the promise of the ALPHA-ML framework in learning polarizabilities of small molecules, future work will focus on the extension of the model to predict $\balpha$ for larger compounds, oligomers and condensed-phase systems.
These developments will allow the design of polarizable force fields for atomistic simulations~\cite{Fanourgakis2008,Lopes2009,pond+10jpcb,medd+14jctc,Cisneros2016} as well as the computational prediction of Raman scattering~\cite{Luber2014,Zhang2016} and sum-frequency generation spectroscopy~\cite{Medders2016,Morita2000,Morita2002},
with better accuracy and lower computational cost,
improving the predictive power of simulations and increasing the insight that can be obtained from experiments.
\section{Methods}
\subsection*{First Principles Calculations of the Dipole Polarizabilities in the QM7b Database}

In this work, DFT calculations with the B3LYP functional and LR-CCSD calculations were conducted using \texttt{Psi4} v1.1~\cite{doi:10.1021/acs.jctc.7b00174}, while DFT calculations with the SCAN0 functional were performed using \texttt{Q-Chem} v5.0~\cite{doi:10.1080/00268976.2014.952696}. 
All of the geometries used for model learning are from the QM7b database~\cite{mont+13njp}.
To be consistent with the optimization process used by the QM7b database, the $52$ showcase molecules (obtained from the ChemSpider database), were optimized using DFT with the PBE functional in a converged numerical basis in \texttt{FHI-AIMS}~\cite{BLUM20092175} (with tight settings/tier 2 basis set).
All geometries were translated to their corresponding center of charge before polarizability calculations.
In all DFT calculations
the finite-field method was implemented, in which the polarizabilites are calculated as the first derivative of the dipole moments with respect to an external electric field. \textit{i.e.} $\balpha=\partial\bmu/\partial \boldsymbol E$,
where $\bmu$ is the dipole moment vector and $\boldsymbol E$ is the external electric field. In practice, a numerical differentiation method was used and the $\delta\boldsymbol E$ added along the $x,y,z$ directions are all $1.8897261250\times10^{-5}~$a.u.

The polarizabilities at CCSD level are calculated directly from linear response coupled-cluster with singles and doubles (LR-CCSD) except for those of 8 large showcase molecules (labelled (18) Phenylalanine, (19) Tyrosine, (20) Tryptophan, (21) Caffeine, (23) Aspirin, (25)  Acyclovir, (26) D-Fructose and (28) D-Glucose in Fig.~\ref{fig:all-show}), which had to be calculated by the (orbital non-relaxed) finite-field method because of the large computational cost in terms of RAM and disk memory. The frozen core approximation and direct \texttt{scf\_type} was used in all CCSD calculations.
In the cases where the finite-field method was used, the polarizabilites of the 8 large showcase molecules were obtained by taking the second derivative of their electric field perturbed energies as $\balpha=\partial^2\boldsymbol{U}/\partial \boldsymbol{E}^2$, where $\boldsymbol{U}$ is the total CCSD energy and the scale of the perturbing field $\delta \boldsymbol E$ is the same as above.

For all calculations, a Dunning-type basis set, d-aug-cc-pVDZ, was used, which was obtained from the EMSL Basis Set Library~\cite{doi:10.1002/(SICI)1096-987X(199610)17:13<1571::AID-JCC9>3.0.CO;2-P,doi:10.1021/ci600510j}. For DFT calculations using \texttt{Q-Chem}, %
a tight threshold with \texttt{scf\_convergence=10} and \texttt{thresh=13} were
assigned for molecules in QM7b Database while \texttt{scf\_convergence=8} combined with \texttt{thresh=14} was used for the showcase molecules. In calculations using \texttt{Psi4}, for the DFT calculations, the convergence criteria for SCF density and energy were both set to $10^{-10}$. For LR-CCSD calculations, all the criteria were used with their default values. The two SCF criteria were set to $5.0\times 10^{-10}$ in the few cases where the SCF calculations did not converge. While dealing with the finite-field CCSD calculations, the two SCF criteria were both set to $5.0\times 10^{-10}$ and a much tighter convergence criterion for the CC amplitude was set, with value of $5.0\times 10^{-9}$, in order to tighten the energy calculations and minimize the possible numerical errors.

\subsection*{Error Assessment}

To assess the accuracy of a given polarizability estimate in a way that includes both scalar and anisotropic components, and is invariant to rigid rotations, we use the Frobenius norm, $\frob{\balpha}^2=\sum_{i,j\in\left\{x,y,z\right\}} \alpha_{ij}^2$.
Based on this metric, and given two sets of polarizabilities, $\balpha_i$ and $\balpha'_i$, for $N$ structures (each of which contains $n_i$ atoms), we define the following statistical error measures. Mean signed error $\text{MSE} \equiv 
\frac{1}{N} \sum_i (\frob{\balpha_i}-\frob{\balpha'_i})/{n_i}$; 
mean absolute error $\text{MAE} \equiv  \frac{1}{N} \sum_i \frob{\balpha_i-\balpha'_i}/{n_i} $; 
root mean square error $\text{RMSE} \equiv  \sqrt{\frac{1}{N} \sum_i {\frob{\balpha_i-\balpha'_i}^2}/{n_i^2}}$.
In these expressions, we have defined each error on a per-atom basis to simplify the comparison between datasets containing molecules of different sizes.

\subsection*{Symmetry-Adapted Gaussian Process Regression}

The SA-GPR framework we use to build a ML model for the polarizability is based on the following steps: (1) Each polarizability tensor $\balpha$ is decomposed into its irreducible (real spherical) components $\alpha^{(0)} = \left(\alpha_{xx}+\alpha_{yy}+\alpha_{zz}\right)/\sqrt{3}$,
which is a scalar, and $
\balpha^{(2)} = \sqrt{2}\left[
\alpha_{xy}, \alpha_{yz}, \alpha_{xz},
\frac{2\alpha_{zz}-\alpha_{xx}-\alpha_{yy}}{2\sqrt{3}},
\frac{\alpha_{xx}-\alpha_{yy}}{2}
\right]$
which is a 5-vector. 
Note that the transformation between Cartesian and spherical components is unitary, so that the Frobenius norm of $\balpha$ is unchanged if computed as $\frob{\balpha}^2=\left|{\alpha^{(0)}}\right|^2+\left|{\balpha^{(2)}}\right|^2$. One can then compute the RMSE error separately on the scalar and vector components of the polarizability.
(2) $\lambda$-SOAP descriptors are computed for each atom-centered environment $\xj$ in the training set; each environment contains information on the interatomic correlations within a prescribed cutoff radius $r_{\rm c}$ around the central atom; we use the notation $\bra{\alpha n l \alpha' n' l'}\ket{\calx^{(2)}_{j,\lambda\mu}}$ to indicate the components that correspond to the environment centred around the $j^{\rm th}$ atom, and are suitable to learn tensor components of order $\lambda$. (3) The base kernel between two environments can be defined as 
\begin{equation}\label{eq:combine_power_spectra}
k_{\mu\mu'}^\lambda(\xj,\xk) = \sum_{\left\{J\right\}} \bra{\calx_{j,\lambda\mu}}\ket{J}\bra{J}\ket{\calx_{k,\lambda\mu'}}^\star,
\end{equation}
where we use the shorthand $\left\{J\right\}$ to indicate  a subset of the possible spherical harmonic components of the descriptors $\ket{\alpha n l \alpha' n' l'}$, that are automatically selected with a farthest point sampling procedure~\cite{imba+18jcp}. 
(4) The linear SOAP kernel can describe atomic correlations up to 3-body terms. Many-body correlations can be introduced by  normalizing it, and raising it to an integer power. Care must be taken to preserve the linear nature of the $\lambda$-SOAP kernels, which is crucial to enforce the correct symmetry properties; we use in practice
\begin{equation}
\begin{split}
k_{\mu\mu'}^{\lambda,\zeta}(\xj,\xk)&\leftarrow k_{\mu\mu'}^{\lambda}(\xj,\xk)\ k^{0}_{00}(\xj,\xk)^{\zeta-1};\\
\mathbf{k}^{\lambda}(\xj,\xk)&\leftarrow {\tilde{\mathbf{k}}^{\lambda}(\xj,\xk)}/{
\sqrt{\left\|\tilde{\mathbf{k}}^{\lambda}(\xj,\xj)\right\|_F\left\|\tilde{\mathbf{k}}^{\lambda}(\xk,\xk)\right\|_F}}.
\end{split}\label{eq:nonlinear}
\end{equation}
(5) we determine the weights $w_{k\mu}$ in a kernel ridge regression model, for each separate component of $\balpha$, which correspond to the optimization of the loss
\begin{equation}
\loss =  \sum_{\mu, \CA \in N}  \bigg|\,\alpha^{(\lambda)}_\mu(\CA) -  \sum_{\substack{k \in M \\ j\in \CA}} w_{ k \mu'} k^\lambda_{\mu\mu'}(\xj,\xk)\bigg|^2 + \sigma^2 \left|\mathbf{w}\right|^2,
\end{equation}
where $N$ is the training set, and $M$ a (possibly sparse) set of representative environments used as basis.
\acknowledgments{
The authors thank Felix Musil and Michael Willatt for helpful discussions. D.M.W. and M.C. acknowledge financial support from the European Research Council under the European Union's Horizon 2020 research and innovation programme (grant agreement no. 677013-HBMAP). A.G. acknowledges funding by the MPG-EPFL Center for Molecular Nanoscience and Technology. Y.Y., K.U.L., and R.A.D. acknowledge partial support from Cornell University through start-up funding. This research used resources of the Argonne Leadership Computing Facility at Argonne National Laboratory, which is supported by the Office of Science of the U.S. Department of Energy under Contract No. DE-AC02-06CH11357.
This work was supported by a grant from the
Swiss National Supercomputing Centre (CSCS) under Project
ID s843, and by computer time from the EPFL scientific computing centre.
}


\begin{thebibliography}{58}%
\makeatletter
\providecommand \@ifxundefined [1]{%
 \@ifx{#1\undefined}
}%
\providecommand \@ifnum [1]{%
 \ifnum #1\expandafter \@firstoftwo
 \else \expandafter \@secondoftwo
 \fi
}%
\providecommand \@ifx [1]{%
 \ifx #1\expandafter \@firstoftwo
 \else \expandafter \@secondoftwo
 \fi
}%
\providecommand \natexlab [1]{#1}%
\providecommand \enquote  [1]{``#1''}%
\providecommand \bibnamefont  [1]{#1}%
\providecommand \bibfnamefont [1]{#1}%
\providecommand \citenamefont [1]{#1}%
\providecommand \href@noop [0]{\@secondoftwo}%
\providecommand \href [0]{\begingroup \@sanitize@url \@href}%
\providecommand \@href[1]{\@@startlink{#1}\@@href}%
\providecommand \@@href[1]{\endgroup#1\@@endlink}%
\providecommand \@sanitize@url [0]{\catcode `\\12\catcode `\$12\catcode
  `\&12\catcode `\#12\catcode `\^12\catcode `\_12\catcode `\%12\relax}%
\providecommand \@@startlink[1]{}%
\providecommand \@@endlink[0]{}%
\providecommand \url  [0]{\begingroup\@sanitize@url \@url }%
\providecommand \@url [1]{\endgroup\@href {#1}{\urlprefix }}%
\providecommand \urlprefix  [0]{URL }%
\providecommand \Eprint [0]{\href }%
\providecommand \doibase [0]{http://dx.doi.org/}%
\providecommand \selectlanguage [0]{\@gobble}%
\providecommand \bibinfo  [0]{\@secondoftwo}%
\providecommand \bibfield  [0]{\@secondoftwo}%
\providecommand \translation [1]{[#1]}%
\providecommand \BibitemOpen [0]{}%
\providecommand \bibitemStop [0]{}%
\providecommand \bibitemNoStop [0]{.\EOS\space}%
\providecommand \EOS [0]{\spacefactor3000\relax}%
\providecommand \BibitemShut  [1]{\csname bibitem#1\endcsname}%
\let\auto@bib@innerbib\@empty
\bibitem [{\citenamefont {Engel}\ and\ \citenamefont
  {Dreizler}(2011)}]{engel2011}%
  \BibitemOpen
  \bibfield  {author} {\bibinfo {author} {\bibfnamefont {E.}~\bibnamefont
  {Engel}}\ and\ \bibinfo {author} {\bibfnamefont {R.~M.}\ \bibnamefont
  {Dreizler}},\ }\href@noop {} {\emph {\bibinfo {title} {{Density Functional
  Theory: An Advanced Course}}}}\ (\bibinfo  {publisher} {Springer},\ \bibinfo
  {address} {Berlin, Heidelberg},\ \bibinfo {year} {2011})\BibitemShut
  {NoStop}%
\bibitem [{\citenamefont {Burke}(2012)}]{burke2012}%
  \BibitemOpen
  \bibfield  {author} {\bibinfo {author} {\bibfnamefont {K.}~\bibnamefont
  {Burke}},\ }\href@noop {} {\bibfield  {journal} {\bibinfo  {journal} {J.
  Chem. Phys.}\ }\textbf {\bibinfo {volume} {136}},\ \bibinfo {pages} {150901}
  (\bibinfo {year} {2012})}\BibitemShut {NoStop}%
\bibitem [{\citenamefont {Lejaeghere}\ \emph {et~al.}(2016)\citenamefont
  {Lejaeghere}, \citenamefont {Bihlmayer}, \citenamefont {Bj{\"o}rkman},
  \citenamefont {Blaha}, \citenamefont {Bl{\"u}gel}, \citenamefont {Blum},
  \citenamefont {Caliste}, \citenamefont {Castelli}, \citenamefont {Clark},
  \citenamefont {Dal~Corso}, \citenamefont {de~Gironcoli}, \citenamefont
  {Deutsch}, \citenamefont {Dewhurst}, \citenamefont {Di~Marco}, \citenamefont
  {Draxl}, \citenamefont {Du{\l}ak}, \citenamefont {Eriksson}, \citenamefont
  {Flores-Livas}, \citenamefont {Garrity}, \citenamefont {Genovese},
  \citenamefont {Giannozzi}, \citenamefont {Giantomassi}, \citenamefont
  {Goedecker}, \citenamefont {Gonze}, \citenamefont {Gr{\r a}n{\"a}s},
  \citenamefont {Gross}, \citenamefont {Gulans}, \citenamefont {Gygi},
  \citenamefont {Hamann}, \citenamefont {Hasnip}, \citenamefont {Holzwarth},
  \citenamefont {Iu{\c s}an}, \citenamefont {Jochym}, \citenamefont {Jollet},
  \citenamefont {Jones}, \citenamefont {Kresse}, \citenamefont {Koepernik},
  \citenamefont {K{\"u}{\c c}{\"u}kbenli}, \citenamefont {Kvashnin},
  \citenamefont {Locht}, \citenamefont {Lubeck}, \citenamefont {Marsman},
  \citenamefont {Marzari}, \citenamefont {Nitzsche}, \citenamefont
  {Nordstr{\"o}m}, \citenamefont {Ozaki}, \citenamefont {Paulatto},
  \citenamefont {Pickard}, \citenamefont {Poelmans}, \citenamefont {Probert},
  \citenamefont {Refson}, \citenamefont {Richter}, \citenamefont {Rignanese},
  \citenamefont {Saha}, \citenamefont {Scheffler}, \citenamefont {Schlipf},
  \citenamefont {Schwarz}, \citenamefont {Sharma}, \citenamefont {Tavazza},
  \citenamefont {Thunstr{\"o}m}, \citenamefont {Tkatchenko}, \citenamefont
  {Torrent}, \citenamefont {Vanderbilt}, \citenamefont {van Setten},
  \citenamefont {Van~Speybroeck}, \citenamefont {Wills}, \citenamefont {Yates},
  \citenamefont {Zhang},\ and\ \citenamefont {Cottenier}}]{cottenier2016}%
  \BibitemOpen
  \bibfield  {author} {\bibinfo {author} {\bibfnamefont {K.}~\bibnamefont
  {Lejaeghere}}, \bibinfo {author} {\bibfnamefont {G.}~\bibnamefont
  {Bihlmayer}}, \bibinfo {author} {\bibfnamefont {T.}~\bibnamefont
  {Bj{\"o}rkman}}, \bibinfo {author} {\bibfnamefont {P.}~\bibnamefont {Blaha}},
  \bibinfo {author} {\bibfnamefont {S.}~\bibnamefont {Bl{\"u}gel}}, \bibinfo
  {author} {\bibfnamefont {V.}~\bibnamefont {Blum}}, \bibinfo {author}
  {\bibfnamefont {D.}~\bibnamefont {Caliste}}, \bibinfo {author} {\bibfnamefont
  {I.~E.}\ \bibnamefont {Castelli}}, \bibinfo {author} {\bibfnamefont {S.~J.}\
  \bibnamefont {Clark}}, \bibinfo {author} {\bibfnamefont {A.}~\bibnamefont
  {Dal~Corso}}, \bibinfo {author} {\bibfnamefont {S.}~\bibnamefont
  {de~Gironcoli}}, \bibinfo {author} {\bibfnamefont {T.}~\bibnamefont
  {Deutsch}}, \bibinfo {author} {\bibfnamefont {J.~K.}\ \bibnamefont
  {Dewhurst}}, \bibinfo {author} {\bibfnamefont {I.}~\bibnamefont {Di~Marco}},
  \bibinfo {author} {\bibfnamefont {C.}~\bibnamefont {Draxl}}, \bibinfo
  {author} {\bibfnamefont {M.}~\bibnamefont {Du{\l}ak}}, \bibinfo {author}
  {\bibfnamefont {O.}~\bibnamefont {Eriksson}}, \bibinfo {author}
  {\bibfnamefont {J.~A.}\ \bibnamefont {Flores-Livas}}, \bibinfo {author}
  {\bibfnamefont {K.~F.}\ \bibnamefont {Garrity}}, \bibinfo {author}
  {\bibfnamefont {L.}~\bibnamefont {Genovese}}, \bibinfo {author}
  {\bibfnamefont {P.}~\bibnamefont {Giannozzi}}, \bibinfo {author}
  {\bibfnamefont {M.}~\bibnamefont {Giantomassi}}, \bibinfo {author}
  {\bibfnamefont {S.}~\bibnamefont {Goedecker}}, \bibinfo {author}
  {\bibfnamefont {X.}~\bibnamefont {Gonze}}, \bibinfo {author} {\bibfnamefont
  {O.}~\bibnamefont {Gr{\r a}n{\"a}s}}, \bibinfo {author} {\bibfnamefont
  {E.~K.~U.}\ \bibnamefont {Gross}}, \bibinfo {author} {\bibfnamefont
  {A.}~\bibnamefont {Gulans}}, \bibinfo {author} {\bibfnamefont
  {F.}~\bibnamefont {Gygi}}, \bibinfo {author} {\bibfnamefont {D.~R.}\
  \bibnamefont {Hamann}}, \bibinfo {author} {\bibfnamefont {P.~J.}\
  \bibnamefont {Hasnip}}, \bibinfo {author} {\bibfnamefont {N.~A.~W.}\
  \bibnamefont {Holzwarth}}, \bibinfo {author} {\bibfnamefont {D.}~\bibnamefont
  {Iu{\c s}an}}, \bibinfo {author} {\bibfnamefont {D.~B.}\ \bibnamefont
  {Jochym}}, \bibinfo {author} {\bibfnamefont {F.}~\bibnamefont {Jollet}},
  \bibinfo {author} {\bibfnamefont {D.}~\bibnamefont {Jones}}, \bibinfo
  {author} {\bibfnamefont {G.}~\bibnamefont {Kresse}}, \bibinfo {author}
  {\bibfnamefont {K.}~\bibnamefont {Koepernik}}, \bibinfo {author}
  {\bibfnamefont {E.}~\bibnamefont {K{\"u}{\c c}{\"u}kbenli}}, \bibinfo
  {author} {\bibfnamefont {Y.~O.}\ \bibnamefont {Kvashnin}}, \bibinfo {author}
  {\bibfnamefont {I.~L.~M.}\ \bibnamefont {Locht}}, \bibinfo {author}
  {\bibfnamefont {S.}~\bibnamefont {Lubeck}}, \bibinfo {author} {\bibfnamefont
  {M.}~\bibnamefont {Marsman}}, \bibinfo {author} {\bibfnamefont
  {N.}~\bibnamefont {Marzari}}, \bibinfo {author} {\bibfnamefont
  {U.}~\bibnamefont {Nitzsche}}, \bibinfo {author} {\bibfnamefont
  {L.}~\bibnamefont {Nordstr{\"o}m}}, \bibinfo {author} {\bibfnamefont
  {T.}~\bibnamefont {Ozaki}}, \bibinfo {author} {\bibfnamefont
  {L.}~\bibnamefont {Paulatto}}, \bibinfo {author} {\bibfnamefont {C.~J.}\
  \bibnamefont {Pickard}}, \bibinfo {author} {\bibfnamefont {W.}~\bibnamefont
  {Poelmans}}, \bibinfo {author} {\bibfnamefont {M.~I.~J.}\ \bibnamefont
  {Probert}}, \bibinfo {author} {\bibfnamefont {K.}~\bibnamefont {Refson}},
  \bibinfo {author} {\bibfnamefont {M.}~\bibnamefont {Richter}}, \bibinfo
  {author} {\bibfnamefont {G.-M.}\ \bibnamefont {Rignanese}}, \bibinfo {author}
  {\bibfnamefont {S.}~\bibnamefont {Saha}}, \bibinfo {author} {\bibfnamefont
  {M.}~\bibnamefont {Scheffler}}, \bibinfo {author} {\bibfnamefont
  {M.}~\bibnamefont {Schlipf}}, \bibinfo {author} {\bibfnamefont
  {K.}~\bibnamefont {Schwarz}}, \bibinfo {author} {\bibfnamefont
  {S.}~\bibnamefont {Sharma}}, \bibinfo {author} {\bibfnamefont
  {F.}~\bibnamefont {Tavazza}}, \bibinfo {author} {\bibfnamefont
  {P.}~\bibnamefont {Thunstr{\"o}m}}, \bibinfo {author} {\bibfnamefont
  {A.}~\bibnamefont {Tkatchenko}}, \bibinfo {author} {\bibfnamefont
  {M.}~\bibnamefont {Torrent}}, \bibinfo {author} {\bibfnamefont
  {D.}~\bibnamefont {Vanderbilt}}, \bibinfo {author} {\bibfnamefont {M.~J.}\
  \bibnamefont {van Setten}}, \bibinfo {author} {\bibfnamefont
  {V.}~\bibnamefont {Van~Speybroeck}}, \bibinfo {author} {\bibfnamefont
  {J.~M.}\ \bibnamefont {Wills}}, \bibinfo {author} {\bibfnamefont {J.~R.}\
  \bibnamefont {Yates}}, \bibinfo {author} {\bibfnamefont {G.-X.}\ \bibnamefont
  {Zhang}}, \ and\ \bibinfo {author} {\bibfnamefont {S.}~\bibnamefont
  {Cottenier}},\ }\href@noop {} {\bibfield  {journal} {\bibinfo  {journal}
  {Science}\ }\textbf {\bibinfo {volume} {351}},\ \bibinfo {pages} {145}
  (\bibinfo {year} {2016})}\BibitemShut {NoStop}%
\bibitem [{\citenamefont {Stone}(1997)}]{stone1997theory}%
  \BibitemOpen
  \bibfield  {author} {\bibinfo {author} {\bibfnamefont {A.}~\bibnamefont
  {Stone}},\ }\href {https://books.google.com/books?id=nQvhXDVnmowC} {\emph
  {\bibinfo {title} {The Theory of Intermolecular Forces}}},\ International
  Series of Monographs on Chemistry\ (\bibinfo  {publisher} {Clarendon Press},\
  \bibinfo {year} {1997})\BibitemShut {NoStop}%
\bibitem [{\citenamefont {Hermann}\ \emph {et~al.}(2017)\citenamefont
  {Hermann}, \citenamefont {DiStasio},\ and\ \citenamefont
  {Tkatchenko}}]{doi:10.1021/acs.chemrev.6b00446}%
  \BibitemOpen
  \bibfield  {author} {\bibinfo {author} {\bibfnamefont {J.}~\bibnamefont
  {Hermann}}, \bibinfo {author} {\bibfnamefont {R.~A.}\ \bibnamefont
  {DiStasio}}, \ and\ \bibinfo {author} {\bibfnamefont {A.}~\bibnamefont
  {Tkatchenko}},\ }\href {\doibase 10.1021/acs.chemrev.6b00446} {\bibfield
  {journal} {\bibinfo  {journal} {Chem. Rev.}\ }\textbf {\bibinfo {volume}
  {117}},\ \bibinfo {pages} {4714} (\bibinfo {year} {2017})}\BibitemShut
  {NoStop}%
\bibitem [{\citenamefont {Tkatchenko}\ and\ \citenamefont
  {Scheffler}(2009)}]{tkat-sche09prl}%
  \BibitemOpen
  \bibfield  {author} {\bibinfo {author} {\bibfnamefont {A.}~\bibnamefont
  {Tkatchenko}}\ and\ \bibinfo {author} {\bibfnamefont {M.}~\bibnamefont
  {Scheffler}},\ }\href@noop {} {\bibfield  {journal} {\bibinfo  {journal}
  {Phys. Rev. Lett.}\ }\textbf {\bibinfo {volume} {102}},\ \bibinfo {pages}
  {073005} (\bibinfo {year} {2009})}\BibitemShut {NoStop}%
\bibitem [{\citenamefont {Tkatchenko}\ \emph {et~al.}(2012)\citenamefont
  {Tkatchenko}, \citenamefont {DiStasio}, \citenamefont {Car},\ and\
  \citenamefont {Scheffler}}]{tkatchenko2012}%
  \BibitemOpen
  \bibfield  {author} {\bibinfo {author} {\bibfnamefont {A.}~\bibnamefont
  {Tkatchenko}}, \bibinfo {author} {\bibfnamefont {R.~A.}\ \bibnamefont
  {DiStasio}}, \bibinfo {author} {\bibfnamefont {R.}~\bibnamefont {Car}}, \
  and\ \bibinfo {author} {\bibfnamefont {M.}~\bibnamefont {Scheffler}},\
  }\href@noop {} {\bibfield  {journal} {\bibinfo  {journal} {Phys. Rev. Lett.}\
  }\textbf {\bibinfo {volume} {108}},\ \bibinfo {pages} {236402} (\bibinfo
  {year} {2012})}\BibitemShut {NoStop}%
\bibitem [{\citenamefont {Grimme}(2011)}]{grimme2011}%
  \BibitemOpen
  \bibfield  {author} {\bibinfo {author} {\bibfnamefont {S.}~\bibnamefont
  {Grimme}},\ }\href@noop {} {\bibfield  {journal} {\bibinfo  {journal} {WIREs
  Computat. Mol. Sci.}\ }\textbf {\bibinfo {volume} {1}},\ \bibinfo {pages}
  {211} (\bibinfo {year} {2011})}\BibitemShut {NoStop}%
\bibitem [{\citenamefont {Grimme}(2014)}]{grimme_chapter}%
  \BibitemOpen
  \bibfield  {author} {\bibinfo {author} {\bibfnamefont {S.}~\bibnamefont
  {Grimme}},\ }\enquote {\bibinfo {title} {Dispersion interaction and chemical
  bonding},}\ in\ \href@noop {} {\emph {\bibinfo {booktitle} {The Chemical
  Bond}}}\ (\bibinfo  {publisher} {Wiley-Blackwell},\ \bibinfo {year} {2014})\
  Chap.~\bibinfo {chapter} {16}, pp.\ \bibinfo {pages} {477--500}\BibitemShut
  {NoStop}%
\bibitem [{\citenamefont {Luber}\ \emph {et~al.}(2014)\citenamefont {Luber},
  \citenamefont {Iannuzzi},\ and\ \citenamefont {Hutter}}]{Luber2014}%
  \BibitemOpen
  \bibfield  {author} {\bibinfo {author} {\bibfnamefont {S.}~\bibnamefont
  {Luber}}, \bibinfo {author} {\bibfnamefont {M.}~\bibnamefont {Iannuzzi}}, \
  and\ \bibinfo {author} {\bibfnamefont {J.}~\bibnamefont {Hutter}},\
  }\href@noop {} {\bibfield  {journal} {\bibinfo  {journal} {J. Chem. Phys.}\
  }\textbf {\bibinfo {volume} {141}},\ \bibinfo {pages} {094503} (\bibinfo
  {year} {2014})}\BibitemShut {NoStop}%
\bibitem [{\citenamefont {Zhang}\ \emph {et~al.}(2016)\citenamefont {Zhang},
  \citenamefont {Tan}, \citenamefont {Wu}, \citenamefont {Shi},\ and\
  \citenamefont {Tan}}]{Zhang2016}%
  \BibitemOpen
  \bibfield  {author} {\bibinfo {author} {\bibfnamefont {X.}~\bibnamefont
  {Zhang}}, \bibinfo {author} {\bibfnamefont {Q.-H.}\ \bibnamefont {Tan}},
  \bibinfo {author} {\bibfnamefont {J.-B.}\ \bibnamefont {Wu}}, \bibinfo
  {author} {\bibfnamefont {W.}~\bibnamefont {Shi}}, \ and\ \bibinfo {author}
  {\bibfnamefont {P.-H.}\ \bibnamefont {Tan}},\ }\href@noop {} {\bibfield
  {journal} {\bibinfo  {journal} {Nanoscale}\ }\textbf {\bibinfo {volume}
  {8}},\ \bibinfo {pages} {6435} (\bibinfo {year} {2016})}\BibitemShut
  {NoStop}%
\bibitem [{\citenamefont {Medders}\ and\ \citenamefont
  {Paesani}(2016)}]{Medders2016}%
  \BibitemOpen
  \bibfield  {author} {\bibinfo {author} {\bibfnamefont {G.~R.}\ \bibnamefont
  {Medders}}\ and\ \bibinfo {author} {\bibfnamefont {F.}~\bibnamefont
  {Paesani}},\ }\href@noop {} {\bibfield  {journal} {\bibinfo  {journal} {Chem.
  Phys. Lett.}\ }\textbf {\bibinfo {volume} {138}},\ \bibinfo {pages} {11}
  (\bibinfo {year} {2016})}\BibitemShut {NoStop}%
\bibitem [{\citenamefont {Morita}\ and\ \citenamefont
  {Hynes}(2000)}]{Morita2000}%
  \BibitemOpen
  \bibfield  {author} {\bibinfo {author} {\bibfnamefont {A.}~\bibnamefont
  {Morita}}\ and\ \bibinfo {author} {\bibfnamefont {J.~T.}\ \bibnamefont
  {Hynes}},\ }\href@noop {} {\bibfield  {journal} {\bibinfo  {journal} {Chem.
  Phys.}\ }\textbf {\bibinfo {volume} {258}},\ \bibinfo {pages} {371} (\bibinfo
  {year} {2000})}\BibitemShut {NoStop}%
\bibitem [{\citenamefont {Morita}\ and\ \citenamefont
  {Hynes}(2002)}]{Morita2002}%
  \BibitemOpen
  \bibfield  {author} {\bibinfo {author} {\bibfnamefont {A.}~\bibnamefont
  {Morita}}\ and\ \bibinfo {author} {\bibfnamefont {J.~T.}\ \bibnamefont
  {Hynes}},\ }\href@noop {} {\bibfield  {journal} {\bibinfo  {journal} {J.
  Phys. Chem. B}\ }\textbf {\bibinfo {volume} {106}},\ \bibinfo {pages} {673}
  (\bibinfo {year} {2002})}\BibitemShut {NoStop}%
\bibitem [{\citenamefont {Shen}(1989)}]{shen1989}%
  \BibitemOpen
  \bibfield  {author} {\bibinfo {author} {\bibfnamefont {Y.~R.}\ \bibnamefont
  {Shen}},\ }\href@noop {} {\bibfield  {journal} {\bibinfo  {journal} {Nature}\
  }\textbf {\bibinfo {volume} {337}},\ \bibinfo {pages} {519} (\bibinfo {year}
  {1989})}\BibitemShut {NoStop}%
\bibitem [{\citenamefont {Sprik}\ and\ \citenamefont
  {Klein}(1988)}]{Sprik1988}%
  \BibitemOpen
  \bibfield  {author} {\bibinfo {author} {\bibfnamefont {M.}~\bibnamefont
  {Sprik}}\ and\ \bibinfo {author} {\bibfnamefont {M.~L.}\ \bibnamefont
  {Klein}},\ }\href {\doibase 10.1063/1.455722} {\bibfield  {journal} {\bibinfo
   {journal} {The Journal of Chemical Physics}\ }\textbf {\bibinfo {volume}
  {89}},\ \bibinfo {pages} {7556} (\bibinfo {year} {1988})}\BibitemShut
  {NoStop}%
\bibitem [{\citenamefont {Fanourgakis}\ and\ \citenamefont
  {Xantheas}(2008)}]{Fanourgakis2008}%
  \BibitemOpen
  \bibfield  {author} {\bibinfo {author} {\bibfnamefont {G.~S.}\ \bibnamefont
  {Fanourgakis}}\ and\ \bibinfo {author} {\bibfnamefont {S.~S.}\ \bibnamefont
  {Xantheas}},\ }\href@noop {} {\bibfield  {journal} {\bibinfo  {journal} {J.
  Chem. Phys.}\ }\textbf {\bibinfo {volume} {128}},\ \bibinfo {pages} {074506}
  (\bibinfo {year} {2008})}\BibitemShut {NoStop}%
\bibitem [{\citenamefont {Lopes}\ \emph {et~al.}(2009)\citenamefont {Lopes},
  \citenamefont {Roux},\ and\ \citenamefont {MacKerell}}]{Lopes2009}%
  \BibitemOpen
  \bibfield  {author} {\bibinfo {author} {\bibfnamefont {P.~E.~M.}\
  \bibnamefont {Lopes}}, \bibinfo {author} {\bibfnamefont {B.}~\bibnamefont
  {Roux}}, \ and\ \bibinfo {author} {\bibfnamefont {A.~D.}\ \bibnamefont
  {MacKerell}},\ }\href@noop {} {\bibfield  {journal} {\bibinfo  {journal}
  {Theor. Chem. Acc.}\ }\textbf {\bibinfo {volume} {124}},\ \bibinfo {pages}
  {11} (\bibinfo {year} {2009})}\BibitemShut {NoStop}%
\bibitem [{\citenamefont {Ponder}\ \emph {et~al.}(2010)\citenamefont {Ponder},
  \citenamefont {Wu}, \citenamefont {Ren}, \citenamefont {Pande}, \citenamefont
  {Chodera}, \citenamefont {Schnieders}, \citenamefont {Haque}, \citenamefont
  {Mobley}, \citenamefont {Lambrecht}, \citenamefont {DiStasio}, \citenamefont
  {Head-Gordon}, \citenamefont {Clark}, \citenamefont {Johnson},\ and\
  \citenamefont {Head-Gordon}}]{pond+10jpcb}%
  \BibitemOpen
  \bibfield  {author} {\bibinfo {author} {\bibfnamefont {J.~W.}\ \bibnamefont
  {Ponder}}, \bibinfo {author} {\bibfnamefont {C.}~\bibnamefont {Wu}}, \bibinfo
  {author} {\bibfnamefont {P.}~\bibnamefont {Ren}}, \bibinfo {author}
  {\bibfnamefont {V.~S.}\ \bibnamefont {Pande}}, \bibinfo {author}
  {\bibfnamefont {J.~D.}\ \bibnamefont {Chodera}}, \bibinfo {author}
  {\bibfnamefont {M.~J.}\ \bibnamefont {Schnieders}}, \bibinfo {author}
  {\bibfnamefont {I.}~\bibnamefont {Haque}}, \bibinfo {author} {\bibfnamefont
  {D.~L.}\ \bibnamefont {Mobley}}, \bibinfo {author} {\bibfnamefont {D.~S.}\
  \bibnamefont {Lambrecht}}, \bibinfo {author} {\bibfnamefont {R.~A.}\
  \bibnamefont {DiStasio}}, \bibinfo {author} {\bibfnamefont {M.}~\bibnamefont
  {Head-Gordon}}, \bibinfo {author} {\bibfnamefont {G.~N.~I.}\ \bibnamefont
  {Clark}}, \bibinfo {author} {\bibfnamefont {M.~E.}\ \bibnamefont {Johnson}},
  \ and\ \bibinfo {author} {\bibfnamefont {T.}~\bibnamefont {Head-Gordon}},\
  }\href@noop {} {\bibfield  {journal} {\bibinfo  {journal} {J. Phys. Chem. B}\
  }\textbf {\bibinfo {volume} {114}},\ \bibinfo {pages} {2549} (\bibinfo {year}
  {2010})}\BibitemShut {NoStop}%
\bibitem [{\citenamefont {Medders}\ \emph {et~al.}(2014)\citenamefont
  {Medders}, \citenamefont {Babin},\ and\ \citenamefont
  {Paesani}}]{medd+14jctc}%
  \BibitemOpen
  \bibfield  {author} {\bibinfo {author} {\bibfnamefont {G.~R.}\ \bibnamefont
  {Medders}}, \bibinfo {author} {\bibfnamefont {V.}~\bibnamefont {Babin}}, \
  and\ \bibinfo {author} {\bibfnamefont {F.}~\bibnamefont {Paesani}},\
  }\href@noop {} {\bibfield  {journal} {\bibinfo  {journal} {J. Chem. Theory
  Comput.}\ }\textbf {\bibinfo {volume} {10}},\ \bibinfo {pages} {2906}
  (\bibinfo {year} {2014})}\BibitemShut {NoStop}%
\bibitem [{\citenamefont {Cisneros}\ \emph {et~al.}(2016)\citenamefont
  {Cisneros}, \citenamefont {Wikfeldt}, \citenamefont {Ojam{\"a}e},
  \citenamefont {Lu}, \citenamefont {Xu}, \citenamefont {Torabifard},
  \citenamefont {Bart{\'o}k}, \citenamefont {Cs{\'a}nyi}, \citenamefont
  {Molinero},\ and\ \citenamefont {Paesani}}]{Cisneros2016}%
  \BibitemOpen
  \bibfield  {author} {\bibinfo {author} {\bibfnamefont {G.~A.}\ \bibnamefont
  {Cisneros}}, \bibinfo {author} {\bibfnamefont {K.~T.}\ \bibnamefont
  {Wikfeldt}}, \bibinfo {author} {\bibfnamefont {L.}~\bibnamefont
  {Ojam{\"a}e}}, \bibinfo {author} {\bibfnamefont {J.}~\bibnamefont {Lu}},
  \bibinfo {author} {\bibfnamefont {Y.}~\bibnamefont {Xu}}, \bibinfo {author}
  {\bibfnamefont {H.}~\bibnamefont {Torabifard}}, \bibinfo {author}
  {\bibfnamefont {A.~P.}\ \bibnamefont {Bart{\'o}k}}, \bibinfo {author}
  {\bibfnamefont {G.}~\bibnamefont {Cs{\'a}nyi}}, \bibinfo {author}
  {\bibfnamefont {V.}~\bibnamefont {Molinero}}, \ and\ \bibinfo {author}
  {\bibfnamefont {F.}~\bibnamefont {Paesani}},\ }\href@noop {} {\bibfield
  {journal} {\bibinfo  {journal} {Chem. Rev.}\ }\textbf {\bibinfo {volume}
  {116}},\ \bibinfo {pages} {7501} (\bibinfo {year} {2016})}\BibitemShut
  {NoStop}%
\bibitem [{\citenamefont {Hait}\ and\ \citenamefont
  {Head-Gordon}(2018)}]{hait_head-gordon_2018}%
  \BibitemOpen
  \bibfield  {author} {\bibinfo {author} {\bibfnamefont {D.}~\bibnamefont
  {Hait}}\ and\ \bibinfo {author} {\bibfnamefont {M.}~\bibnamefont
  {Head-Gordon}},\ }\href@noop {} {\bibfield  {journal} {\bibinfo  {journal}
  {Phys. Chem. Chem. Phys.}\ }\textbf {\bibinfo {volume} {20}},\ \bibinfo
  {pages} {19800} (\bibinfo {year} {2018})}\BibitemShut {NoStop}%
\bibitem [{\citenamefont {Hammond}\ \emph {et~al.}(2008)\citenamefont
  {Hammond}, \citenamefont {de~Jong},\ and\ \citenamefont
  {Kowalski}}]{doi:10.1063/1.2929840}%
  \BibitemOpen
  \bibfield  {author} {\bibinfo {author} {\bibfnamefont {J.~R.}\ \bibnamefont
  {Hammond}}, \bibinfo {author} {\bibfnamefont {W.~A.}\ \bibnamefont
  {de~Jong}}, \ and\ \bibinfo {author} {\bibfnamefont {K.}~\bibnamefont
  {Kowalski}},\ }\href@noop {} {\bibfield  {journal} {\bibinfo  {journal} {J.
  Chem. Phys.}\ }\textbf {\bibinfo {volume} {128}},\ \bibinfo {pages} {224102}
  (\bibinfo {year} {2008})}\BibitemShut {NoStop}%
\bibitem [{\citenamefont {Lao}\ \emph {et~al.}(2018{\natexlab{a}})\citenamefont
  {Lao}, \citenamefont {Jia}, \citenamefont {Gladue}, \citenamefont
  {Csernica},\ and\ \citenamefont {{DiStasio Jr.}}}]{LaoJiaGla18}%
  \BibitemOpen
  \bibfield  {author} {\bibinfo {author} {\bibfnamefont {K.~U.}\ \bibnamefont
  {Lao}}, \bibinfo {author} {\bibfnamefont {J.}~\bibnamefont {Jia}}, \bibinfo
  {author} {\bibfnamefont {C.}~\bibnamefont {Gladue}}, \bibinfo {author}
  {\bibfnamefont {P.}~\bibnamefont {Csernica}}, \ and\ \bibinfo {author}
  {\bibfnamefont {R.~A.}\ \bibnamefont {{DiStasio Jr.}}},\ }\href@noop {} {\
  (\bibinfo {year} {2018}{\natexlab{a}})},\ \bibinfo {note} {(to be
  submitted)}\BibitemShut {NoStop}%
\bibitem [{\citenamefont {Behler}\ and\ \citenamefont
  {Parrinello}(2007)}]{behl-parr07prl}%
  \BibitemOpen
  \bibfield  {author} {\bibinfo {author} {\bibfnamefont {J.}~\bibnamefont
  {Behler}}\ and\ \bibinfo {author} {\bibfnamefont {M.}~\bibnamefont
  {Parrinello}},\ }\href@noop {} {\bibfield  {journal} {\bibinfo  {journal}
  {Phys. Rev. Lett.}\ }\textbf {\bibinfo {volume} {98}},\ \bibinfo {pages}
  {146401} (\bibinfo {year} {2007})}\BibitemShut {NoStop}%
\bibitem [{\citenamefont {Bart{\'{o}}k}\ \emph {et~al.}(2010)\citenamefont
  {Bart{\'{o}}k}, \citenamefont {Payne}, \citenamefont {Kondor},\ and\
  \citenamefont {Cs{\'{a}}nyi}}]{bart+10prl}%
  \BibitemOpen
  \bibfield  {author} {\bibinfo {author} {\bibfnamefont {A.~P.}\ \bibnamefont
  {Bart{\'{o}}k}}, \bibinfo {author} {\bibfnamefont {M.~C.}\ \bibnamefont
  {Payne}}, \bibinfo {author} {\bibfnamefont {R.}~\bibnamefont {Kondor}}, \
  and\ \bibinfo {author} {\bibfnamefont {G.}~\bibnamefont {Cs{\'{a}}nyi}},\
  }\href@noop {} {\bibfield  {journal} {\bibinfo  {journal} {Phys. Rev. Lett.}\
  }\textbf {\bibinfo {volume} {104}},\ \bibinfo {pages} {136403} (\bibinfo
  {year} {2010})}\BibitemShut {NoStop}%
\bibitem [{\citenamefont {Rupp}\ \emph {et~al.}(2012)\citenamefont {Rupp},
  \citenamefont {Tkatchenko}, \citenamefont {M{\"{u}}ller},\ and\ \citenamefont
  {von Lilienfeld}}]{rupp+12prl}%
  \BibitemOpen
  \bibfield  {author} {\bibinfo {author} {\bibfnamefont {M.}~\bibnamefont
  {Rupp}}, \bibinfo {author} {\bibfnamefont {A.}~\bibnamefont {Tkatchenko}},
  \bibinfo {author} {\bibfnamefont {K.-R.}\ \bibnamefont {M{\"{u}}ller}}, \
  and\ \bibinfo {author} {\bibfnamefont {O.~A.}\ \bibnamefont {von
  Lilienfeld}},\ }\href@noop {} {\bibfield  {journal} {\bibinfo  {journal}
  {Phys. Rev. Lett.}\ }\textbf {\bibinfo {volume} {108}},\ \bibinfo {pages}
  {058301} (\bibinfo {year} {2012})}\BibitemShut {NoStop}%
\bibitem [{\citenamefont {De}\ \emph {et~al.}(2016)\citenamefont {De},
  \citenamefont {Bart{\'{o}}k}, \citenamefont {Cs{\'{a}}nyi},\ and\
  \citenamefont {Ceriotti}}]{de+16pccp}%
  \BibitemOpen
  \bibfield  {author} {\bibinfo {author} {\bibfnamefont {S.}~\bibnamefont
  {De}}, \bibinfo {author} {\bibfnamefont {A.~P.}\ \bibnamefont
  {Bart{\'{o}}k}}, \bibinfo {author} {\bibfnamefont {G.}~\bibnamefont
  {Cs{\'{a}}nyi}}, \ and\ \bibinfo {author} {\bibfnamefont {M.}~\bibnamefont
  {Ceriotti}},\ }\href@noop {} {\bibfield  {journal} {\bibinfo  {journal}
  {Phys. Chem. Chem. Phys.}\ }\textbf {\bibinfo {volume} {18}},\ \bibinfo
  {pages} {13754} (\bibinfo {year} {2016})}\BibitemShut {NoStop}%
\bibitem [{\citenamefont {Faber}\ \emph {et~al.}(2017)\citenamefont {Faber},
  \citenamefont {Hutchison}, \citenamefont {Huang}, \citenamefont {Gilmer},
  \citenamefont {Schoenholz}, \citenamefont {Dahl}, \citenamefont {Vinyals},
  \citenamefont {Kearnes}, \citenamefont {Riley},\ and\ \citenamefont {von
  Lilienfeld}}]{fabe+17jctc}%
  \BibitemOpen
  \bibfield  {author} {\bibinfo {author} {\bibfnamefont {F.~A.}\ \bibnamefont
  {Faber}}, \bibinfo {author} {\bibfnamefont {L.}~\bibnamefont {Hutchison}},
  \bibinfo {author} {\bibfnamefont {B.}~\bibnamefont {Huang}}, \bibinfo
  {author} {\bibfnamefont {J.}~\bibnamefont {Gilmer}}, \bibinfo {author}
  {\bibfnamefont {S.~S.}\ \bibnamefont {Schoenholz}}, \bibinfo {author}
  {\bibfnamefont {G.~E.}\ \bibnamefont {Dahl}}, \bibinfo {author}
  {\bibfnamefont {O.}~\bibnamefont {Vinyals}}, \bibinfo {author} {\bibfnamefont
  {S.}~\bibnamefont {Kearnes}}, \bibinfo {author} {\bibfnamefont {P.~F.}\
  \bibnamefont {Riley}}, \ and\ \bibinfo {author} {\bibfnamefont {O.~A.}\
  \bibnamefont {von Lilienfeld}},\ }\href@noop {} {\bibfield  {journal}
  {\bibinfo  {journal} {J. Chem. Theory Comput.}\ }\textbf {\bibinfo {volume}
  {13}},\ \bibinfo {pages} {5255} (\bibinfo {year} {2017})}\BibitemShut
  {NoStop}%
\bibitem [{\citenamefont {Ramakrishnan}\ \emph {et~al.}(2015)\citenamefont
  {Ramakrishnan}, \citenamefont {Dral}, \citenamefont {Rupp},\ and\
  \citenamefont {von Lilienfeld}}]{rama+15jctc}%
  \BibitemOpen
  \bibfield  {author} {\bibinfo {author} {\bibfnamefont {R.}~\bibnamefont
  {Ramakrishnan}}, \bibinfo {author} {\bibfnamefont {P.~O.}\ \bibnamefont
  {Dral}}, \bibinfo {author} {\bibfnamefont {M.}~\bibnamefont {Rupp}}, \ and\
  \bibinfo {author} {\bibfnamefont {O.~A.}\ \bibnamefont {von Lilienfeld}},\
  }\href@noop {} {\bibfield  {journal} {\bibinfo  {journal} {J. Chem. Theory
  Comput.}\ }\textbf {\bibinfo {volume} {11}},\ \bibinfo {pages} {2087}
  (\bibinfo {year} {2015})}\BibitemShut {NoStop}%
\bibitem [{\citenamefont {Bart{\'{o}}k}\ \emph {et~al.}(2017)\citenamefont
  {Bart{\'{o}}k}, \citenamefont {De}, \citenamefont {Poelking}, \citenamefont
  {Bernstein}, \citenamefont {Kermode}, \citenamefont {Cs{\'{a}}nyi},\ and\
  \citenamefont {Ceriotti}}]{bart+17sa}%
  \BibitemOpen
  \bibfield  {author} {\bibinfo {author} {\bibfnamefont {A.~P.}\ \bibnamefont
  {Bart{\'{o}}k}}, \bibinfo {author} {\bibfnamefont {S.}~\bibnamefont {De}},
  \bibinfo {author} {\bibfnamefont {C.}~\bibnamefont {Poelking}}, \bibinfo
  {author} {\bibfnamefont {N.}~\bibnamefont {Bernstein}}, \bibinfo {author}
  {\bibfnamefont {J.~R.}\ \bibnamefont {Kermode}}, \bibinfo {author}
  {\bibfnamefont {G.}~\bibnamefont {Cs{\'{a}}nyi}}, \ and\ \bibinfo {author}
  {\bibfnamefont {M.}~\bibnamefont {Ceriotti}},\ }\href@noop {} {\bibfield
  {journal} {\bibinfo  {journal} {Sci. Adv.}\ }\textbf {\bibinfo {volume}
  {3}},\ \bibinfo {pages} {e1701816} (\bibinfo {year} {2017})}\BibitemShut
  {NoStop}%
\bibitem [{\citenamefont {Bereau}\ \emph {et~al.}(2015)\citenamefont {Bereau},
  \citenamefont {Andrienko},\ and\ \citenamefont {von
  Lilienfeld}}]{tris+15jctc}%
  \BibitemOpen
  \bibfield  {author} {\bibinfo {author} {\bibfnamefont {T.}~\bibnamefont
  {Bereau}}, \bibinfo {author} {\bibfnamefont {D.}~\bibnamefont {Andrienko}}, \
  and\ \bibinfo {author} {\bibfnamefont {O.~A.}\ \bibnamefont {von
  Lilienfeld}},\ }\href@noop {} {\bibfield  {journal} {\bibinfo  {journal} {J.
  Chem. Theory Comput.}\ }\textbf {\bibinfo {volume} {11}},\ \bibinfo {pages}
  {3225} (\bibinfo {year} {2015})}\BibitemShut {NoStop}%
\bibitem [{\citenamefont {Liang}\ \emph {et~al.}(2017)\citenamefont {Liang},
  \citenamefont {Tocci}, \citenamefont {Wilkins}, \citenamefont {Grisafi},
  \citenamefont {Roke},\ and\ \citenamefont {Ceriotti}}]{lian+17prb}%
  \BibitemOpen
  \bibfield  {author} {\bibinfo {author} {\bibfnamefont {C.}~\bibnamefont
  {Liang}}, \bibinfo {author} {\bibfnamefont {G.}~\bibnamefont {Tocci}},
  \bibinfo {author} {\bibfnamefont {D.~M.}\ \bibnamefont {Wilkins}}, \bibinfo
  {author} {\bibfnamefont {A.}~\bibnamefont {Grisafi}}, \bibinfo {author}
  {\bibfnamefont {S.}~\bibnamefont {Roke}}, \ and\ \bibinfo {author}
  {\bibfnamefont {M.}~\bibnamefont {Ceriotti}},\ }\href@noop {} {\bibfield
  {journal} {\bibinfo  {journal} {Phys. Rev. B}\ }\textbf {\bibinfo {volume}
  {96}},\ \bibinfo {pages} {041407} (\bibinfo {year} {2017})}\BibitemShut
  {NoStop}%
\bibitem [{\citenamefont {Grisafi}\ \emph {et~al.}(2018)\citenamefont
  {Grisafi}, \citenamefont {Wilkins}, \citenamefont {Cs{\'{a}}nyi},\ and\
  \citenamefont {Ceriotti}}]{gris+18prl}%
  \BibitemOpen
  \bibfield  {author} {\bibinfo {author} {\bibfnamefont {A.}~\bibnamefont
  {Grisafi}}, \bibinfo {author} {\bibfnamefont {D.~M.}\ \bibnamefont
  {Wilkins}}, \bibinfo {author} {\bibfnamefont {G.}~\bibnamefont
  {Cs{\'{a}}nyi}}, \ and\ \bibinfo {author} {\bibfnamefont {M.}~\bibnamefont
  {Ceriotti}},\ }\href@noop {} {\bibfield  {journal} {\bibinfo  {journal}
  {Phys. Rev. Lett.}\ }\textbf {\bibinfo {volume} {120}},\ \bibinfo {pages}
  {036002} (\bibinfo {year} {2018})}\BibitemShut {NoStop}%
\bibitem [{\citenamefont {Montavon}\ \emph {et~al.}(2013)\citenamefont
  {Montavon}, \citenamefont {Rupp}, \citenamefont {Gobre}, \citenamefont
  {Vazquez-Mayagoitia}, \citenamefont {Hansen}, \citenamefont {Tkatchenko},
  \citenamefont {M{\"{u}}ller},\ and\ \citenamefont {{Anatole Von
  Lilienfeld}}}]{mont+13njp}%
  \BibitemOpen
  \bibfield  {author} {\bibinfo {author} {\bibfnamefont {G.}~\bibnamefont
  {Montavon}}, \bibinfo {author} {\bibfnamefont {M.}~\bibnamefont {Rupp}},
  \bibinfo {author} {\bibfnamefont {V.}~\bibnamefont {Gobre}}, \bibinfo
  {author} {\bibfnamefont {A.}~\bibnamefont {Vazquez-Mayagoitia}}, \bibinfo
  {author} {\bibfnamefont {K.}~\bibnamefont {Hansen}}, \bibinfo {author}
  {\bibfnamefont {A.}~\bibnamefont {Tkatchenko}}, \bibinfo {author}
  {\bibfnamefont {K.~R.}\ \bibnamefont {M{\"{u}}ller}}, \ and\ \bibinfo
  {author} {\bibfnamefont {O.}~\bibnamefont {{Anatole Von Lilienfeld}}},\
  }\href@noop {} {\bibfield  {journal} {\bibinfo  {journal} {New J. Phys.}\
  }\textbf {\bibinfo {volume} {15}},\ \bibinfo {pages} {095003} (\bibinfo
  {year} {2013})}\BibitemShut {NoStop}%
\bibitem [{\citenamefont {Blum}\ and\ \citenamefont {Reymond}(2009)}]{blum}%
  \BibitemOpen
  \bibfield  {author} {\bibinfo {author} {\bibfnamefont {L.~C.}\ \bibnamefont
  {Blum}}\ and\ \bibinfo {author} {\bibfnamefont {J.-L.}\ \bibnamefont
  {Reymond}},\ }\href@noop {} {\bibfield  {journal} {\bibinfo  {journal} {J.
  Am. Chem. Soc.}\ }\textbf {\bibinfo {volume} {131}},\ \bibinfo {pages} {8732}
  (\bibinfo {year} {2009})}\BibitemShut {NoStop}%
\bibitem [{\citenamefont {Becke}(1993)}]{doi:10.1063/1.464913}%
  \BibitemOpen
  \bibfield  {author} {\bibinfo {author} {\bibfnamefont {A.~D.}\ \bibnamefont
  {Becke}},\ }\href@noop {} {\bibfield  {journal} {\bibinfo  {journal} {J.
  Chem. Phys.}\ }\textbf {\bibinfo {volume} {98}},\ \bibinfo {pages} {5648}
  (\bibinfo {year} {1993})}\BibitemShut {NoStop}%
\bibitem [{\citenamefont {Stephens}\ \emph {et~al.}(1994)\citenamefont
  {Stephens}, \citenamefont {Devlin}, \citenamefont {Chabalowski},\ and\
  \citenamefont {Frisch}}]{doi:10.1021/j100096a001}%
  \BibitemOpen
  \bibfield  {author} {\bibinfo {author} {\bibfnamefont {P.~J.}\ \bibnamefont
  {Stephens}}, \bibinfo {author} {\bibfnamefont {F.~J.}\ \bibnamefont
  {Devlin}}, \bibinfo {author} {\bibfnamefont {C.~F.}\ \bibnamefont
  {Chabalowski}}, \ and\ \bibinfo {author} {\bibfnamefont {M.~J.}\ \bibnamefont
  {Frisch}},\ }\href@noop {} {\bibfield  {journal} {\bibinfo  {journal} {J.
  Phys. Chem.}\ }\textbf {\bibinfo {volume} {98}},\ \bibinfo {pages} {11623}
  (\bibinfo {year} {1994})}\BibitemShut {NoStop}%
\bibitem [{\citenamefont {Hui}\ and\ \citenamefont
  {Chai}(2016)}]{doi:10.1063/1.4940734}%
  \BibitemOpen
  \bibfield  {author} {\bibinfo {author} {\bibfnamefont {K.}~\bibnamefont
  {Hui}}\ and\ \bibinfo {author} {\bibfnamefont {J.-D.}\ \bibnamefont {Chai}},\
  }\href@noop {} {\bibfield  {journal} {\bibinfo  {journal} {J. Chem. Phys.}\
  }\textbf {\bibinfo {volume} {144}},\ \bibinfo {pages} {044114} (\bibinfo
  {year} {2016})}\BibitemShut {NoStop}%
\bibitem [{\citenamefont {Koch}\ and\ \citenamefont
  {Jørgensen}(1990)}]{doi:10.1063/1.458814}%
  \BibitemOpen
  \bibfield  {author} {\bibinfo {author} {\bibfnamefont {H.}~\bibnamefont
  {Koch}}\ and\ \bibinfo {author} {\bibfnamefont {P.}~\bibnamefont
  {Jørgensen}},\ }\href@noop {} {\bibfield  {journal} {\bibinfo  {journal} {J.
  Chem. Phys.}\ }\textbf {\bibinfo {volume} {93}},\ \bibinfo {pages} {3333}
  (\bibinfo {year} {1990})}\BibitemShut {NoStop}%
\bibitem [{\citenamefont {Ove}\ \emph {et~al.}(1998)\citenamefont {Ove},
  \citenamefont {Poul},\ and\ \citenamefont
  {Christof}}]{doi:10.1002/(SICI)1097-461X(1998)68:1<1::AID-QUA1>3.0.CO;2-Z}%
  \BibitemOpen
  \bibfield  {author} {\bibinfo {author} {\bibfnamefont {C.}~\bibnamefont
  {Ove}}, \bibinfo {author} {\bibfnamefont {J.}~\bibnamefont {Poul}}, \ and\
  \bibinfo {author} {\bibfnamefont {H.}~\bibnamefont {Christof}},\ }\href@noop
  {} {\bibfield  {journal} {\bibinfo  {journal} {Int. J. Quantum Chem.}\
  }\textbf {\bibinfo {volume} {68}},\ \bibinfo {pages} {1} (\bibinfo {year}
  {1998})}\BibitemShut {NoStop}%
\bibitem [{\citenamefont {Hammond}\ \emph {et~al.}(2009)\citenamefont
  {Hammond}, \citenamefont {Govind}, \citenamefont {Kowalski}, \citenamefont
  {Autschbach},\ and\ \citenamefont {Xantheas}}]{HamGovKow09}%
  \BibitemOpen
  \bibfield  {author} {\bibinfo {author} {\bibfnamefont {J.~R.}\ \bibnamefont
  {Hammond}}, \bibinfo {author} {\bibfnamefont {N.}~\bibnamefont {Govind}},
  \bibinfo {author} {\bibfnamefont {K.}~\bibnamefont {Kowalski}}, \bibinfo
  {author} {\bibfnamefont {J.}~\bibnamefont {Autschbach}}, \ and\ \bibinfo
  {author} {\bibfnamefont {S.~S.}\ \bibnamefont {Xantheas}},\ }\href@noop {}
  {\bibfield  {journal} {\bibinfo  {journal} {J. Chem. Phys.}\ }\textbf
  {\bibinfo {volume} {131}},\ \bibinfo {pages} {214103:1} (\bibinfo {year}
  {2009})}\BibitemShut {NoStop}%
\bibitem [{\citenamefont {Lao}\ \emph {et~al.}(2018{\natexlab{b}})\citenamefont
  {Lao}, \citenamefont {Jia}, \citenamefont {Maitra},\ and\ \citenamefont
  {{DiStasio Jr.}}}]{LaoJiaMai18}%
  \BibitemOpen
  \bibfield  {author} {\bibinfo {author} {\bibfnamefont {K.~U.}\ \bibnamefont
  {Lao}}, \bibinfo {author} {\bibfnamefont {J.}~\bibnamefont {Jia}}, \bibinfo
  {author} {\bibfnamefont {R.}~\bibnamefont {Maitra}}, \ and\ \bibinfo {author}
  {\bibfnamefont {R.~A.}\ \bibnamefont {{DiStasio Jr.}}},\ }\href@noop {}
  {\bibfield  {journal} {\bibinfo  {journal} {J. Chem. Phys.}\ } (\bibinfo
  {year} {2018}{\natexlab{b}})},\ \bibinfo {note} {(in revision)}\BibitemShut
  {NoStop}%
\bibitem [{\citenamefont {Kendall}\ \emph {et~al.}(1992)\citenamefont
  {Kendall}, \citenamefont {Dunning},\ and\ \citenamefont
  {Harrison}}]{doi:10.1063/1.462569}%
  \BibitemOpen
  \bibfield  {author} {\bibinfo {author} {\bibfnamefont {R.~A.}\ \bibnamefont
  {Kendall}}, \bibinfo {author} {\bibfnamefont {T.~H.}\ \bibnamefont
  {Dunning}}, \ and\ \bibinfo {author} {\bibfnamefont {R.~J.}\ \bibnamefont
  {Harrison}},\ }\href@noop {} {\bibfield  {journal} {\bibinfo  {journal} {J.
  Chem. Phys.}\ }\textbf {\bibinfo {volume} {96}},\ \bibinfo {pages} {6796}
  (\bibinfo {year} {1992})}\BibitemShut {NoStop}%
\bibitem [{\citenamefont {Christiansen}\ \emph {et~al.}(1999)\citenamefont
  {Christiansen}, \citenamefont {Gauss},\ and\ \citenamefont
  {Stanton}}]{ChrGauSta99}%
  \BibitemOpen
  \bibfield  {author} {\bibinfo {author} {\bibfnamefont {O.}~\bibnamefont
  {Christiansen}}, \bibinfo {author} {\bibfnamefont {J.}~\bibnamefont {Gauss}},
  \ and\ \bibinfo {author} {\bibfnamefont {J.~F.}\ \bibnamefont {Stanton}},\
  }\href@noop {} {\bibfield  {journal} {\bibinfo  {journal} {Chem. Phys.
  Lett.}\ }\textbf {\bibinfo {volume} {305}},\ \bibinfo {pages} {147} (\bibinfo
  {year} {1999})}\BibitemShut {NoStop}%
\bibitem [{\citenamefont {Karne}\ \emph {et~al.}(2015)\citenamefont {Karne},
  \citenamefont {Vaval}, \citenamefont {Pal}, \citenamefont {Vásquez-Pérez},
  \citenamefont {Köster},\ and\ \citenamefont {Calaminici}}]{KARNE2015168}%
  \BibitemOpen
  \bibfield  {author} {\bibinfo {author} {\bibfnamefont {A.~S.}\ \bibnamefont
  {Karne}}, \bibinfo {author} {\bibfnamefont {N.}~\bibnamefont {Vaval}},
  \bibinfo {author} {\bibfnamefont {S.}~\bibnamefont {Pal}}, \bibinfo {author}
  {\bibfnamefont {J.~M.}\ \bibnamefont {Vásquez-Pérez}}, \bibinfo {author}
  {\bibfnamefont {A.~M.}\ \bibnamefont {Köster}}, \ and\ \bibinfo {author}
  {\bibfnamefont {P.}~\bibnamefont {Calaminici}},\ }\href@noop {} {\bibfield
  {journal} {\bibinfo  {journal} {Chem. Phys. Lett.}\ }\textbf {\bibinfo
  {volume} {635}},\ \bibinfo {pages} {168} (\bibinfo {year}
  {2015})}\BibitemShut {NoStop}%
\bibitem [{\citenamefont {Imbalzano}\ \emph {et~al.}(2018)\citenamefont
  {Imbalzano}, \citenamefont {Anelli}, \citenamefont {Giofr{\'{e}}},
  \citenamefont {Klees}, \citenamefont {Behler},\ and\ \citenamefont
  {Ceriotti}}]{imba+18jcp}%
  \BibitemOpen
  \bibfield  {author} {\bibinfo {author} {\bibfnamefont {G.}~\bibnamefont
  {Imbalzano}}, \bibinfo {author} {\bibfnamefont {A.}~\bibnamefont {Anelli}},
  \bibinfo {author} {\bibfnamefont {D.}~\bibnamefont {Giofr{\'{e}}}}, \bibinfo
  {author} {\bibfnamefont {S.}~\bibnamefont {Klees}}, \bibinfo {author}
  {\bibfnamefont {J.}~\bibnamefont {Behler}}, \ and\ \bibinfo {author}
  {\bibfnamefont {M.}~\bibnamefont {Ceriotti}},\ }\href@noop {} {\bibfield
  {journal} {\bibinfo  {journal} {J. Chem. Phys.}\ }\textbf {\bibinfo {volume}
  {148}},\ \bibinfo {pages} {241730} (\bibinfo {year} {2018})}\BibitemShut
  {NoStop}%
\bibitem [{\citenamefont {Bart{\'{o}}k}\ \emph {et~al.}(2013)\citenamefont
  {Bart{\'{o}}k}, \citenamefont {Kondor},\ and\ \citenamefont
  {Cs{\'{a}}nyi}}]{bart+13prb}%
  \BibitemOpen
  \bibfield  {author} {\bibinfo {author} {\bibfnamefont {A.~P.}\ \bibnamefont
  {Bart{\'{o}}k}}, \bibinfo {author} {\bibfnamefont {R.}~\bibnamefont
  {Kondor}}, \ and\ \bibinfo {author} {\bibfnamefont {G.}~\bibnamefont
  {Cs{\'{a}}nyi}},\ }\href@noop {} {\bibfield  {journal} {\bibinfo  {journal}
  {Phys. Rev. B}\ }\textbf {\bibinfo {volume} {87}},\ \bibinfo {pages} {184115}
  (\bibinfo {year} {2013})}\BibitemShut {NoStop}%
\bibitem [{\citenamefont {Glielmo}\ \emph {et~al.}(2018)\citenamefont
  {Glielmo}, \citenamefont {Zeni},\ and\ \citenamefont {{De
  Vita}}}]{glie+18prb}%
  \BibitemOpen
  \bibfield  {author} {\bibinfo {author} {\bibfnamefont {A.}~\bibnamefont
  {Glielmo}}, \bibinfo {author} {\bibfnamefont {C.}~\bibnamefont {Zeni}}, \
  and\ \bibinfo {author} {\bibfnamefont {A.}~\bibnamefont {{De Vita}}},\
  }\href@noop {} {\bibfield  {journal} {\bibinfo  {journal} {Phys. Rev. B}\
  }\textbf {\bibinfo {volume} {97}},\ \bibinfo {pages} {184307} (\bibinfo
  {year} {2018})}\BibitemShut {NoStop}%
\bibitem [{\citenamefont {Willatt}\ \emph {et~al.}(2018)\citenamefont
  {Willatt}, \citenamefont {Musil},\ and\ \citenamefont
  {Ceriotti}}]{density-arxiv}%
  \BibitemOpen
  \bibfield  {author} {\bibinfo {author} {\bibfnamefont {M.~J.}\ \bibnamefont
  {Willatt}}, \bibinfo {author} {\bibfnamefont {F.}~\bibnamefont {Musil}}, \
  and\ \bibinfo {author} {\bibfnamefont {M.}~\bibnamefont {Ceriotti}},\
  }\href@noop {} {\  (\bibinfo {year} {2018})},\ \bibinfo {note}
  {arXiv:1807.00408. Preprint, posted on Jul 1, 2018}\BibitemShut {NoStop}%
\bibitem [{\citenamefont {Laidig}\ and\ \citenamefont
  {Bader}(1990)}]{laidig1990}%
  \BibitemOpen
  \bibfield  {author} {\bibinfo {author} {\bibfnamefont {K.~E.}\ \bibnamefont
  {Laidig}}\ and\ \bibinfo {author} {\bibfnamefont {R.~F.~W.}\ \bibnamefont
  {Bader}},\ }\href@noop {} {\bibfield  {journal} {\bibinfo  {journal} {J.
  Chem. Phys.}\ }\textbf {\bibinfo {volume} {93}},\ \bibinfo {pages} {7213}
  (\bibinfo {year} {1990})}\BibitemShut {NoStop}%
\bibitem [{\citenamefont {Applequist}\ \emph {et~al.}(1972)\citenamefont
  {Applequist}, \citenamefont {Carl},\ and\ \citenamefont
  {Fung}}]{applequist1972}%
  \BibitemOpen
  \bibfield  {author} {\bibinfo {author} {\bibfnamefont {J.}~\bibnamefont
  {Applequist}}, \bibinfo {author} {\bibfnamefont {J.~R.}\ \bibnamefont
  {Carl}}, \ and\ \bibinfo {author} {\bibfnamefont {K.-K.}\ \bibnamefont
  {Fung}},\ }\href@noop {} {\bibfield  {journal} {\bibinfo  {journal} {J. Am.
  Chem. Soc.}\ }\textbf {\bibinfo {volume} {94}},\ \bibinfo {pages} {2952}
  (\bibinfo {year} {1972})}\BibitemShut {NoStop}%
\bibitem [{\citenamefont {DelloStritto}\ and\ \citenamefont
  {Sofo}(2017)}]{DelloStritto2017}%
  \BibitemOpen
  \bibfield  {author} {\bibinfo {author} {\bibfnamefont {M.}~\bibnamefont
  {DelloStritto}}\ and\ \bibinfo {author} {\bibfnamefont {J.}~\bibnamefont
  {Sofo}},\ }\href@noop {} {\bibfield  {journal} {\bibinfo  {journal} {J. Phys.
  Chem. A}\ }\textbf {\bibinfo {volume} {121}},\ \bibinfo {pages} {3045}
  (\bibinfo {year} {2017})}\BibitemShut {NoStop}%
\bibitem [{\citenamefont {Parrish}\ \emph {et~al.}(2017)\citenamefont
  {Parrish}, \citenamefont {Burns}, \citenamefont {Smith}, \citenamefont
  {Simmonett}, \citenamefont {DePrince}, \citenamefont {Hohenstein},
  \citenamefont {Bozkaya}, \citenamefont {Sokolov}, \citenamefont {Di~Remigio},
  \citenamefont {Richard}, \citenamefont {Gonthier}, \citenamefont {James},
  \citenamefont {McAlexander}, \citenamefont {Kumar}, \citenamefont {Saitow},
  \citenamefont {Wang}, \citenamefont {Pritchard}, \citenamefont {Verma},
  \citenamefont {Schaefer}, \citenamefont {Patkowski}, \citenamefont {King},
  \citenamefont {Valeev}, \citenamefont {Evangelista}, \citenamefont {Turney},
  \citenamefont {Crawford},\ and\ \citenamefont
  {Sherrill}}]{doi:10.1021/acs.jctc.7b00174}%
  \BibitemOpen
  \bibfield  {author} {\bibinfo {author} {\bibfnamefont {R.~M.}\ \bibnamefont
  {Parrish}}, \bibinfo {author} {\bibfnamefont {L.~A.}\ \bibnamefont {Burns}},
  \bibinfo {author} {\bibfnamefont {D.~G.~A.}\ \bibnamefont {Smith}}, \bibinfo
  {author} {\bibfnamefont {A.~C.}\ \bibnamefont {Simmonett}}, \bibinfo {author}
  {\bibfnamefont {A.~E.}\ \bibnamefont {DePrince}}, \bibinfo {author}
  {\bibfnamefont {E.~G.}\ \bibnamefont {Hohenstein}}, \bibinfo {author}
  {\bibfnamefont {U.}~\bibnamefont {Bozkaya}}, \bibinfo {author} {\bibfnamefont
  {A.~Y.}\ \bibnamefont {Sokolov}}, \bibinfo {author} {\bibfnamefont
  {R.}~\bibnamefont {Di~Remigio}}, \bibinfo {author} {\bibfnamefont {R.~M.}\
  \bibnamefont {Richard}}, \bibinfo {author} {\bibfnamefont {J.~F.}\
  \bibnamefont {Gonthier}}, \bibinfo {author} {\bibfnamefont {A.~M.}\
  \bibnamefont {James}}, \bibinfo {author} {\bibfnamefont {H.~R.}\ \bibnamefont
  {McAlexander}}, \bibinfo {author} {\bibfnamefont {A.}~\bibnamefont {Kumar}},
  \bibinfo {author} {\bibfnamefont {M.}~\bibnamefont {Saitow}}, \bibinfo
  {author} {\bibfnamefont {X.}~\bibnamefont {Wang}}, \bibinfo {author}
  {\bibfnamefont {B.~P.}\ \bibnamefont {Pritchard}}, \bibinfo {author}
  {\bibfnamefont {P.}~\bibnamefont {Verma}}, \bibinfo {author} {\bibfnamefont
  {H.~F.}\ \bibnamefont {Schaefer}}, \bibinfo {author} {\bibfnamefont
  {K.}~\bibnamefont {Patkowski}}, \bibinfo {author} {\bibfnamefont {R.~A.}\
  \bibnamefont {King}}, \bibinfo {author} {\bibfnamefont {E.~F.}\ \bibnamefont
  {Valeev}}, \bibinfo {author} {\bibfnamefont {F.~A.}\ \bibnamefont
  {Evangelista}}, \bibinfo {author} {\bibfnamefont {J.~M.}\ \bibnamefont
  {Turney}}, \bibinfo {author} {\bibfnamefont {T.~D.}\ \bibnamefont
  {Crawford}}, \ and\ \bibinfo {author} {\bibfnamefont {C.~D.}\ \bibnamefont
  {Sherrill}},\ }\href@noop {} {\bibfield  {journal} {\bibinfo  {journal} {J.
  Chem. Theory Comput.}\ }\textbf {\bibinfo {volume} {13}},\ \bibinfo {pages}
  {3185} (\bibinfo {year} {2017})}\BibitemShut {NoStop}%
\bibitem [{\citenamefont {Shao}\ \emph {et~al.}(2015)\citenamefont {Shao},
  \citenamefont {Gan}, \citenamefont {Epifanovsky}, \citenamefont {Gilbert},
  \citenamefont {Wormit}, \citenamefont {Kussmann}, \citenamefont {Lange},
  \citenamefont {Behn}, \citenamefont {Deng}, \citenamefont {Feng},
  \citenamefont {Ghosh}, \citenamefont {Goldey}, \citenamefont {Horn},
  \citenamefont {Jacobson}, \citenamefont {Kaliman}, \citenamefont
  {Khaliullin}, \citenamefont {Kuś}, \citenamefont {Landau}, \citenamefont
  {Liu}, \citenamefont {Proynov}, \citenamefont {Rhee}, \citenamefont
  {Richard}, \citenamefont {Rohrdanz}, \citenamefont {Steele}, \citenamefont
  {Sundstrom}, \citenamefont {III}, \citenamefont {Zimmerman}, \citenamefont
  {Zuev}, \citenamefont {Albrecht}, \citenamefont {Alguire}, \citenamefont
  {Austin}, \citenamefont {Beran}, \citenamefont {Bernard}, \citenamefont
  {Berquist}, \citenamefont {Brandhorst}, \citenamefont {Bravaya},
  \citenamefont {Brown}, \citenamefont {Casanova}, \citenamefont {Chang},
  \citenamefont {Chen}, \citenamefont {Chien}, \citenamefont {Closser},
  \citenamefont {Crittenden}, \citenamefont {Diedenhofen}, \citenamefont {Jr.},
  \citenamefont {Do}, \citenamefont {Dutoi}, \citenamefont {Edgar},
  \citenamefont {Fatehi}, \citenamefont {Fusti-Molnar}, \citenamefont
  {Ghysels}, \citenamefont {Golubeva-Zadorozhnaya}, \citenamefont {Gomes},
  \citenamefont {Hanson-Heine}, \citenamefont {Harbach}, \citenamefont
  {Hauser}, \citenamefont {Hohenstein}, \citenamefont {Holden}, \citenamefont
  {Jagau}, \citenamefont {Ji}, \citenamefont {Kaduk}, \citenamefont
  {Khistyaev}, \citenamefont {Kim}, \citenamefont {Kim}, \citenamefont {King},
  \citenamefont {Klunzinger}, \citenamefont {Kosenkov}, \citenamefont
  {Kowalczyk}, \citenamefont {Krauter}, \citenamefont {Lao}, \citenamefont
  {Laurent}, \citenamefont {Lawler}, \citenamefont {Levchenko}, \citenamefont
  {Lin}, \citenamefont {Liu}, \citenamefont {Livshits}, \citenamefont {Lochan},
  \citenamefont {Luenser}, \citenamefont {Manohar}, \citenamefont {Manzer},
  \citenamefont {Mao}, \citenamefont {Mardirossian}, \citenamefont {Marenich},
  \citenamefont {Maurer}, \citenamefont {Mayhall}, \citenamefont {Neuscamman},
  \citenamefont {Oana}, \citenamefont {Olivares-Amaya}, \citenamefont
  {O’Neill}, \citenamefont {Parkhill}, \citenamefont {Perrine}, \citenamefont
  {Peverati}, \citenamefont {Prociuk}, \citenamefont {Rehn}, \citenamefont
  {Rosta}, \citenamefont {Russ}, \citenamefont {Sharada}, \citenamefont
  {Sharma}, \citenamefont {Small}, \citenamefont {Sodt}, \citenamefont {Stein},
  \citenamefont {Stück}, \citenamefont {Su}, \citenamefont {Thom},
  \citenamefont {Tsuchimochi}, \citenamefont {Vanovschi}, \citenamefont {Vogt},
  \citenamefont {Vydrov}, \citenamefont {Wang}, \citenamefont {Watson},
  \citenamefont {Wenzel}, \citenamefont {White}, \citenamefont {Williams},
  \citenamefont {Yang}, \citenamefont {Yeganeh}, \citenamefont {Yost},
  \citenamefont {You}, \citenamefont {Zhang}, \citenamefont {Zhang},
  \citenamefont {Zhao}, \citenamefont {Brooks}, \citenamefont {Chan},
  \citenamefont {Chipman}, \citenamefont {Cramer}, \citenamefont {III},
  \citenamefont {Gordon}, \citenamefont {Hehre}, \citenamefont {Klamt},
  \citenamefont {III}, \citenamefont {Schmidt}, \citenamefont {Sherrill},
  \citenamefont {Truhlar}, \citenamefont {Warshel}, \citenamefont {Xu},
  \citenamefont {Aspuru-Guzik}, \citenamefont {Baer}, \citenamefont {Bell},
  \citenamefont {Besley}, \citenamefont {Chai}, \citenamefont {Dreuw},
  \citenamefont {Dunietz}, \citenamefont {Furlani}, \citenamefont {Gwaltney},
  \citenamefont {Hsu}, \citenamefont {Jung}, \citenamefont {Kong},
  \citenamefont {Lambrecht}, \citenamefont {Liang}, \citenamefont {Ochsenfeld},
  \citenamefont {Rassolov}, \citenamefont {Slipchenko}, \citenamefont
  {Subotnik}, \citenamefont {Voorhis}, \citenamefont {Herbert}, \citenamefont
  {Krylov}, \citenamefont {Gill},\ and\ \citenamefont
  {Head-Gordon}}]{doi:10.1080/00268976.2014.952696}%
  \BibitemOpen
  \bibfield  {author} {\bibinfo {author} {\bibfnamefont {Y.}~\bibnamefont
  {Shao}}, \bibinfo {author} {\bibfnamefont {Z.}~\bibnamefont {Gan}}, \bibinfo
  {author} {\bibfnamefont {E.}~\bibnamefont {Epifanovsky}}, \bibinfo {author}
  {\bibfnamefont {A.~T.}\ \bibnamefont {Gilbert}}, \bibinfo {author}
  {\bibfnamefont {M.}~\bibnamefont {Wormit}}, \bibinfo {author} {\bibfnamefont
  {J.}~\bibnamefont {Kussmann}}, \bibinfo {author} {\bibfnamefont {A.~W.}\
  \bibnamefont {Lange}}, \bibinfo {author} {\bibfnamefont {A.}~\bibnamefont
  {Behn}}, \bibinfo {author} {\bibfnamefont {J.}~\bibnamefont {Deng}}, \bibinfo
  {author} {\bibfnamefont {X.}~\bibnamefont {Feng}}, \bibinfo {author}
  {\bibfnamefont {D.}~\bibnamefont {Ghosh}}, \bibinfo {author} {\bibfnamefont
  {M.}~\bibnamefont {Goldey}}, \bibinfo {author} {\bibfnamefont {P.~R.}\
  \bibnamefont {Horn}}, \bibinfo {author} {\bibfnamefont {L.~D.}\ \bibnamefont
  {Jacobson}}, \bibinfo {author} {\bibfnamefont {I.}~\bibnamefont {Kaliman}},
  \bibinfo {author} {\bibfnamefont {R.~Z.}\ \bibnamefont {Khaliullin}},
  \bibinfo {author} {\bibfnamefont {T.}~\bibnamefont {Kuś}}, \bibinfo {author}
  {\bibfnamefont {A.}~\bibnamefont {Landau}}, \bibinfo {author} {\bibfnamefont
  {J.}~\bibnamefont {Liu}}, \bibinfo {author} {\bibfnamefont {E.~I.}\
  \bibnamefont {Proynov}}, \bibinfo {author} {\bibfnamefont {Y.~M.}\
  \bibnamefont {Rhee}}, \bibinfo {author} {\bibfnamefont {R.~M.}\ \bibnamefont
  {Richard}}, \bibinfo {author} {\bibfnamefont {M.~A.}\ \bibnamefont
  {Rohrdanz}}, \bibinfo {author} {\bibfnamefont {R.~P.}\ \bibnamefont
  {Steele}}, \bibinfo {author} {\bibfnamefont {E.~J.}\ \bibnamefont
  {Sundstrom}}, \bibinfo {author} {\bibfnamefont {H.~L.~W.}\ \bibnamefont
  {III}}, \bibinfo {author} {\bibfnamefont {P.~M.}\ \bibnamefont {Zimmerman}},
  \bibinfo {author} {\bibfnamefont {D.}~\bibnamefont {Zuev}}, \bibinfo {author}
  {\bibfnamefont {B.}~\bibnamefont {Albrecht}}, \bibinfo {author}
  {\bibfnamefont {E.}~\bibnamefont {Alguire}}, \bibinfo {author} {\bibfnamefont
  {B.}~\bibnamefont {Austin}}, \bibinfo {author} {\bibfnamefont {G.~J.~O.}\
  \bibnamefont {Beran}}, \bibinfo {author} {\bibfnamefont {Y.~A.}\ \bibnamefont
  {Bernard}}, \bibinfo {author} {\bibfnamefont {E.}~\bibnamefont {Berquist}},
  \bibinfo {author} {\bibfnamefont {K.}~\bibnamefont {Brandhorst}}, \bibinfo
  {author} {\bibfnamefont {K.~B.}\ \bibnamefont {Bravaya}}, \bibinfo {author}
  {\bibfnamefont {S.~T.}\ \bibnamefont {Brown}}, \bibinfo {author}
  {\bibfnamefont {D.}~\bibnamefont {Casanova}}, \bibinfo {author}
  {\bibfnamefont {C.-M.}\ \bibnamefont {Chang}}, \bibinfo {author}
  {\bibfnamefont {Y.}~\bibnamefont {Chen}}, \bibinfo {author} {\bibfnamefont
  {S.~H.}\ \bibnamefont {Chien}}, \bibinfo {author} {\bibfnamefont {K.~D.}\
  \bibnamefont {Closser}}, \bibinfo {author} {\bibfnamefont {D.~L.}\
  \bibnamefont {Crittenden}}, \bibinfo {author} {\bibfnamefont
  {M.}~\bibnamefont {Diedenhofen}}, \bibinfo {author} {\bibfnamefont
  {R.~A.~D.}\ \bibnamefont {Jr.}}, \bibinfo {author} {\bibfnamefont
  {H.}~\bibnamefont {Do}}, \bibinfo {author} {\bibfnamefont {A.~D.}\
  \bibnamefont {Dutoi}}, \bibinfo {author} {\bibfnamefont {R.~G.}\ \bibnamefont
  {Edgar}}, \bibinfo {author} {\bibfnamefont {S.}~\bibnamefont {Fatehi}},
  \bibinfo {author} {\bibfnamefont {L.}~\bibnamefont {Fusti-Molnar}}, \bibinfo
  {author} {\bibfnamefont {A.}~\bibnamefont {Ghysels}}, \bibinfo {author}
  {\bibfnamefont {A.}~\bibnamefont {Golubeva-Zadorozhnaya}}, \bibinfo {author}
  {\bibfnamefont {J.}~\bibnamefont {Gomes}}, \bibinfo {author} {\bibfnamefont
  {M.~W.}\ \bibnamefont {Hanson-Heine}}, \bibinfo {author} {\bibfnamefont
  {P.~H.}\ \bibnamefont {Harbach}}, \bibinfo {author} {\bibfnamefont {A.~W.}\
  \bibnamefont {Hauser}}, \bibinfo {author} {\bibfnamefont {E.~G.}\
  \bibnamefont {Hohenstein}}, \bibinfo {author} {\bibfnamefont {Z.~C.}\
  \bibnamefont {Holden}}, \bibinfo {author} {\bibfnamefont {T.-C.}\
  \bibnamefont {Jagau}}, \bibinfo {author} {\bibfnamefont {H.}~\bibnamefont
  {Ji}}, \bibinfo {author} {\bibfnamefont {B.}~\bibnamefont {Kaduk}}, \bibinfo
  {author} {\bibfnamefont {K.}~\bibnamefont {Khistyaev}}, \bibinfo {author}
  {\bibfnamefont {J.}~\bibnamefont {Kim}}, \bibinfo {author} {\bibfnamefont
  {J.}~\bibnamefont {Kim}}, \bibinfo {author} {\bibfnamefont {R.~A.}\
  \bibnamefont {King}}, \bibinfo {author} {\bibfnamefont {P.}~\bibnamefont
  {Klunzinger}}, \bibinfo {author} {\bibfnamefont {D.}~\bibnamefont
  {Kosenkov}}, \bibinfo {author} {\bibfnamefont {T.}~\bibnamefont {Kowalczyk}},
  \bibinfo {author} {\bibfnamefont {C.~M.}\ \bibnamefont {Krauter}}, \bibinfo
  {author} {\bibfnamefont {K.~U.}\ \bibnamefont {Lao}}, \bibinfo {author}
  {\bibfnamefont {A.~D.}\ \bibnamefont {Laurent}}, \bibinfo {author}
  {\bibfnamefont {K.~V.}\ \bibnamefont {Lawler}}, \bibinfo {author}
  {\bibfnamefont {S.~V.}\ \bibnamefont {Levchenko}}, \bibinfo {author}
  {\bibfnamefont {C.~Y.}\ \bibnamefont {Lin}}, \bibinfo {author} {\bibfnamefont
  {F.}~\bibnamefont {Liu}}, \bibinfo {author} {\bibfnamefont {E.}~\bibnamefont
  {Livshits}}, \bibinfo {author} {\bibfnamefont {R.~C.}\ \bibnamefont
  {Lochan}}, \bibinfo {author} {\bibfnamefont {A.}~\bibnamefont {Luenser}},
  \bibinfo {author} {\bibfnamefont {P.}~\bibnamefont {Manohar}}, \bibinfo
  {author} {\bibfnamefont {S.~F.}\ \bibnamefont {Manzer}}, \bibinfo {author}
  {\bibfnamefont {S.-P.}\ \bibnamefont {Mao}}, \bibinfo {author} {\bibfnamefont
  {N.}~\bibnamefont {Mardirossian}}, \bibinfo {author} {\bibfnamefont {A.~V.}\
  \bibnamefont {Marenich}}, \bibinfo {author} {\bibfnamefont {S.~A.}\
  \bibnamefont {Maurer}}, \bibinfo {author} {\bibfnamefont {N.~J.}\
  \bibnamefont {Mayhall}}, \bibinfo {author} {\bibfnamefont {E.}~\bibnamefont
  {Neuscamman}}, \bibinfo {author} {\bibfnamefont {C.~M.}\ \bibnamefont
  {Oana}}, \bibinfo {author} {\bibfnamefont {R.}~\bibnamefont
  {Olivares-Amaya}}, \bibinfo {author} {\bibfnamefont {D.~P.}\ \bibnamefont
  {O’Neill}}, \bibinfo {author} {\bibfnamefont {J.~A.}\ \bibnamefont
  {Parkhill}}, \bibinfo {author} {\bibfnamefont {T.~M.}\ \bibnamefont
  {Perrine}}, \bibinfo {author} {\bibfnamefont {R.}~\bibnamefont {Peverati}},
  \bibinfo {author} {\bibfnamefont {A.}~\bibnamefont {Prociuk}}, \bibinfo
  {author} {\bibfnamefont {D.~R.}\ \bibnamefont {Rehn}}, \bibinfo {author}
  {\bibfnamefont {E.}~\bibnamefont {Rosta}}, \bibinfo {author} {\bibfnamefont
  {N.~J.}\ \bibnamefont {Russ}}, \bibinfo {author} {\bibfnamefont {S.~M.}\
  \bibnamefont {Sharada}}, \bibinfo {author} {\bibfnamefont {S.}~\bibnamefont
  {Sharma}}, \bibinfo {author} {\bibfnamefont {D.~W.}\ \bibnamefont {Small}},
  \bibinfo {author} {\bibfnamefont {A.}~\bibnamefont {Sodt}}, \bibinfo {author}
  {\bibfnamefont {T.}~\bibnamefont {Stein}}, \bibinfo {author} {\bibfnamefont
  {D.}~\bibnamefont {Stück}}, \bibinfo {author} {\bibfnamefont {Y.-C.}\
  \bibnamefont {Su}}, \bibinfo {author} {\bibfnamefont {A.~J.}\ \bibnamefont
  {Thom}}, \bibinfo {author} {\bibfnamefont {T.}~\bibnamefont {Tsuchimochi}},
  \bibinfo {author} {\bibfnamefont {V.}~\bibnamefont {Vanovschi}}, \bibinfo
  {author} {\bibfnamefont {L.}~\bibnamefont {Vogt}}, \bibinfo {author}
  {\bibfnamefont {O.}~\bibnamefont {Vydrov}}, \bibinfo {author} {\bibfnamefont
  {T.}~\bibnamefont {Wang}}, \bibinfo {author} {\bibfnamefont {M.~A.}\
  \bibnamefont {Watson}}, \bibinfo {author} {\bibfnamefont {J.}~\bibnamefont
  {Wenzel}}, \bibinfo {author} {\bibfnamefont {A.}~\bibnamefont {White}},
  \bibinfo {author} {\bibfnamefont {C.~F.}\ \bibnamefont {Williams}}, \bibinfo
  {author} {\bibfnamefont {J.}~\bibnamefont {Yang}}, \bibinfo {author}
  {\bibfnamefont {S.}~\bibnamefont {Yeganeh}}, \bibinfo {author} {\bibfnamefont
  {S.~R.}\ \bibnamefont {Yost}}, \bibinfo {author} {\bibfnamefont {Z.-Q.}\
  \bibnamefont {You}}, \bibinfo {author} {\bibfnamefont {I.~Y.}\ \bibnamefont
  {Zhang}}, \bibinfo {author} {\bibfnamefont {X.}~\bibnamefont {Zhang}},
  \bibinfo {author} {\bibfnamefont {Y.}~\bibnamefont {Zhao}}, \bibinfo {author}
  {\bibfnamefont {B.~R.}\ \bibnamefont {Brooks}}, \bibinfo {author}
  {\bibfnamefont {G.~K.}\ \bibnamefont {Chan}}, \bibinfo {author}
  {\bibfnamefont {D.~M.}\ \bibnamefont {Chipman}}, \bibinfo {author}
  {\bibfnamefont {C.~J.}\ \bibnamefont {Cramer}}, \bibinfo {author}
  {\bibfnamefont {W.~A.~G.}\ \bibnamefont {III}}, \bibinfo {author}
  {\bibfnamefont {M.~S.}\ \bibnamefont {Gordon}}, \bibinfo {author}
  {\bibfnamefont {W.~J.}\ \bibnamefont {Hehre}}, \bibinfo {author}
  {\bibfnamefont {A.}~\bibnamefont {Klamt}}, \bibinfo {author} {\bibfnamefont
  {H.~F.~S.}\ \bibnamefont {III}}, \bibinfo {author} {\bibfnamefont {M.~W.}\
  \bibnamefont {Schmidt}}, \bibinfo {author} {\bibfnamefont {C.~D.}\
  \bibnamefont {Sherrill}}, \bibinfo {author} {\bibfnamefont {D.~G.}\
  \bibnamefont {Truhlar}}, \bibinfo {author} {\bibfnamefont {A.}~\bibnamefont
  {Warshel}}, \bibinfo {author} {\bibfnamefont {X.}~\bibnamefont {Xu}},
  \bibinfo {author} {\bibfnamefont {A.}~\bibnamefont {Aspuru-Guzik}}, \bibinfo
  {author} {\bibfnamefont {R.}~\bibnamefont {Baer}}, \bibinfo {author}
  {\bibfnamefont {A.~T.}\ \bibnamefont {Bell}}, \bibinfo {author}
  {\bibfnamefont {N.~A.}\ \bibnamefont {Besley}}, \bibinfo {author}
  {\bibfnamefont {J.-D.}\ \bibnamefont {Chai}}, \bibinfo {author}
  {\bibfnamefont {A.}~\bibnamefont {Dreuw}}, \bibinfo {author} {\bibfnamefont
  {B.~D.}\ \bibnamefont {Dunietz}}, \bibinfo {author} {\bibfnamefont {T.~R.}\
  \bibnamefont {Furlani}}, \bibinfo {author} {\bibfnamefont {S.~R.}\
  \bibnamefont {Gwaltney}}, \bibinfo {author} {\bibfnamefont {C.-P.}\
  \bibnamefont {Hsu}}, \bibinfo {author} {\bibfnamefont {Y.}~\bibnamefont
  {Jung}}, \bibinfo {author} {\bibfnamefont {J.}~\bibnamefont {Kong}}, \bibinfo
  {author} {\bibfnamefont {D.~S.}\ \bibnamefont {Lambrecht}}, \bibinfo {author}
  {\bibfnamefont {W.}~\bibnamefont {Liang}}, \bibinfo {author} {\bibfnamefont
  {C.}~\bibnamefont {Ochsenfeld}}, \bibinfo {author} {\bibfnamefont {V.~A.}\
  \bibnamefont {Rassolov}}, \bibinfo {author} {\bibfnamefont {L.~V.}\
  \bibnamefont {Slipchenko}}, \bibinfo {author} {\bibfnamefont {J.~E.}\
  \bibnamefont {Subotnik}}, \bibinfo {author} {\bibfnamefont {T.~V.}\
  \bibnamefont {Voorhis}}, \bibinfo {author} {\bibfnamefont {J.~M.}\
  \bibnamefont {Herbert}}, \bibinfo {author} {\bibfnamefont {A.~I.}\
  \bibnamefont {Krylov}}, \bibinfo {author} {\bibfnamefont {P.~M.}\
  \bibnamefont {Gill}}, \ and\ \bibinfo {author} {\bibfnamefont
  {M.}~\bibnamefont {Head-Gordon}},\ }\href@noop {} {\bibfield  {journal}
  {\bibinfo  {journal} {Mol. Phys.}\ }\textbf {\bibinfo {volume} {113}},\
  \bibinfo {pages} {184} (\bibinfo {year} {2015})}\BibitemShut {NoStop}%
\bibitem [{\citenamefont {Blum}\ \emph {et~al.}(2009)\citenamefont {Blum},
  \citenamefont {Gehrke}, \citenamefont {Hanke}, \citenamefont {Havu},
  \citenamefont {Havu}, \citenamefont {Ren}, \citenamefont {Reuter},\ and\
  \citenamefont {Scheffler}}]{BLUM20092175}%
  \BibitemOpen
  \bibfield  {author} {\bibinfo {author} {\bibfnamefont {V.}~\bibnamefont
  {Blum}}, \bibinfo {author} {\bibfnamefont {R.}~\bibnamefont {Gehrke}},
  \bibinfo {author} {\bibfnamefont {F.}~\bibnamefont {Hanke}}, \bibinfo
  {author} {\bibfnamefont {P.}~\bibnamefont {Havu}}, \bibinfo {author}
  {\bibfnamefont {V.}~\bibnamefont {Havu}}, \bibinfo {author} {\bibfnamefont
  {X.}~\bibnamefont {Ren}}, \bibinfo {author} {\bibfnamefont {K.}~\bibnamefont
  {Reuter}}, \ and\ \bibinfo {author} {\bibfnamefont {M.}~\bibnamefont
  {Scheffler}},\ }\href@noop {} {\bibfield  {journal} {\bibinfo  {journal}
  {Comput. Phys. Commun.}\ }\textbf {\bibinfo {volume} {180}},\ \bibinfo
  {pages} {2175 } (\bibinfo {year} {2009})}\BibitemShut {NoStop}%
\bibitem [{\citenamefont
  {David}(1996)}]{doi:10.1002/(SICI)1096-987X(199610)17:13<1571::AID-JCC9>3.0.CO;2-P}%
  \BibitemOpen
  \bibfield  {author} {\bibinfo {author} {\bibfnamefont {F.}~\bibnamefont
  {David}},\ }\href@noop {} {\bibfield  {journal} {\bibinfo  {journal} {J.
  Comput. Chem.}\ }\textbf {\bibinfo {volume} {17}},\ \bibinfo {pages} {1571}
  (\bibinfo {year} {1996})}\BibitemShut {NoStop}%
\bibitem [{\citenamefont {Schuchardt}\ \emph {et~al.}(2007)\citenamefont
  {Schuchardt}, \citenamefont {Didier}, \citenamefont {Elsethagen},
  \citenamefont {Sun}, \citenamefont {Gurumoorthi}, \citenamefont {Chase},
  \citenamefont {Li},\ and\ \citenamefont {Windus}}]{doi:10.1021/ci600510j}%
  \BibitemOpen
  \bibfield  {author} {\bibinfo {author} {\bibfnamefont {K.~L.}\ \bibnamefont
  {Schuchardt}}, \bibinfo {author} {\bibfnamefont {B.~T.}\ \bibnamefont
  {Didier}}, \bibinfo {author} {\bibfnamefont {T.}~\bibnamefont {Elsethagen}},
  \bibinfo {author} {\bibfnamefont {L.}~\bibnamefont {Sun}}, \bibinfo {author}
  {\bibfnamefont {V.}~\bibnamefont {Gurumoorthi}}, \bibinfo {author}
  {\bibfnamefont {J.}~\bibnamefont {Chase}}, \bibinfo {author} {\bibfnamefont
  {J.}~\bibnamefont {Li}}, \ and\ \bibinfo {author} {\bibfnamefont {T.~L.}\
  \bibnamefont {Windus}},\ }\href@noop {} {\bibfield  {journal} {\bibinfo
  {journal} {J. Chem. Inf. Model.}\ }\textbf {\bibinfo {volume} {47}},\
  \bibinfo {pages} {1045} (\bibinfo {year} {2007})}\BibitemShut {NoStop}%
\end{thebibliography}
\end{document}